\documentclass[apj]{emulateapj}

\usepackage{apjfonts}
\usepackage{epsfig}
\usepackage{graphicx}
\usepackage{url}
\usepackage{natbib}
\usepackage{float}
\usepackage{amsmath}
\usepackage[bf, sf, FIGTOPCAP, nooneline, tight]{subfigure}
\usepackage{graphicx}
\usepackage{dcolumn}
\usepackage{bm}
\usepackage[usenames,dvipsnames]{color}
\usepackage{amssymb}
\usepackage{epstopdf}
\usepackage{latexsym}
\usepackage[T1]{fontenc}
\usepackage{aecompl}
\shorttitle{Cosmology from Tomographic AP}
\shortauthors{Li et al.}

\newcommand{\hMsun}{{\ifmmode{h^{-1}{\rm
        {M_{\odot}}}}\else{$h^{-1}{\rm{M_{\odot}}}$~}\fi}}
\newcommand{\hMpc}{{\ifmmode{h^{-1}{\rm Mpc}}\else{$h^{-1}$Mpc }\fi}}

\newbox\tablebox    \newdimen\tablewidth

\def\leaderfil{\leaders\hbox to 5pt{\hss.\hss}\hfil}

\def\endPlancktable{\tablewidth=\columnwidth
    $$\hss\copy\tablebox\hss$$
    \vskip-\lastskip\vskip -2pt}

\def\tablenote#1 #2\par{\begingroup \parindent=0.8em
    \abovedisplayshortskip=0pt\belowdisplayshortskip=0pt
    \noindent
    $$\hss\vbox{\hsize\tablewidth \hangindent=\parindent \hangafter=1 \noindent
    \hbox to \parindent{$^#1$\hss}\strut#2\strut\par}\hss$$
    \endgroup}
\def\doubleline{\vskip 3pt\hrule \vskip 1.5pt \hrule \vskip 5pt}

\begin{document}

\voffset-1.25cm
\title
{The redshift dependence of Alcock-Paczynski effect:
cosmological constraints from the current and next generation observations}

\author{Xiao-Dong Li\altaffilmark{1},
Haitao Miao\altaffilmark{1},
Xin Wang\altaffilmark{1,6},
Xue~Zhang\altaffilmark{2},
Feng Fang\altaffilmark{1},
Xiaolin Luo\altaffilmark{1},
Qing-Guo~Huang\altaffilmark{2,3,4,5},
Miao Li\altaffilmark{1}
}

\email{lixiaod25@mail.sysu.edu.cn}
\email{Corresponding Author: zhangxue@itp.ac.cn}

\altaffiltext{1}{School of Physics and Astronomy, Sun Yat-Sen University, Zhuhai 519082, P. R. China}
\altaffiltext{2}{CAS Key Laboratory of Theoretical Physics, Institute of Theoretical Physics, Chinese Academy of Sciences, Beijing 100190, China}
\altaffiltext{3}{School of Physical Sciences, University of Chinese Academy of Sciences, No. 19A Yuquan Road, Beijing 100049, China}
\altaffiltext{4}{Synergetic Innovation Center for Quantum Effects and Applications, Hunan Normal University, 36 Lushan Lu, 410081, Changsha, China}
\altaffiltext{5}{Center for Gravitation and cosmology, College of Physical Science and Technology, Yangzhou University, 88 South University Ave., 225009, Yangzhou, China}
\altaffiltext{6}{Canadian Institute for Theoretical Astrophysics, 60 St. George St., Toronto, ON, M5H 3H8, Canada}

\label{firstpage}
\begin{abstract}
The tomographic Alcock-Paczynski (AP) test is a robust large-scale structure (LSS) measurement that receives little contamination from the
redshift space distortion (RSD).
It has placed tight cosmological constraints by using small and intermediate clustering scales of the LSS data.
However, previous works have neglected the cross-correlation among different redshift bins,
which could cause the statistical uncertainty being underestimated by $\sim$20\%.
In this work, we further improve this method by including this multi-redshifts full correlation.
We apply it to the SDSS DR12 galaxies sample and find out that,
for $\Lambda$CDM, the combination of AP with the Planck+BAO dataset slightly reduces (within 1-$\sigma$)
$\Omega_m$ to $0.304\pm0.007$ (68.3\% CL).
This then leads to a larger $H_0$ and also mildly affects $\Omega_b h^2$, $n_s$
and the derived parameters $z_*$, $r_*$, $z_{re}$
but not $\tau$, $A_s$ and $\sigma_8$.
For the flat $w$CDM model,
our measurement gives $\Omega_m=0.301\pm 0.010$ and $w=-1.090\pm 0.047$,
where the additional AP measurement reduces the error budget by $\sim 25\%$.
When including more parameters into the analysis,
the AP method also improves the constraints on $\Omega_k$, $\sum m_\mu$, $N_{\rm eff}$ by $20-30\%$.
Early universe parameters such as $dn_s/d{\rm ln}k$ and $r$, however, are unaffected.
Assuming the dark energy equation of state $w=w_0+w_a \frac{z}{1+z}$,
the Planck+BAO+SNIa+$H_0$+AP datasets prefer a dynamical dark energy at $\approx1.5 \sigma$ CL.
Finally, we forecast the cosmological constraints expected from the DESI galaxy survey and
find that combining AP with CMB+BAO method would improve the $w_0$-$w_a$ constraint by a factor of $\sim 10$.
\end{abstract}

\maketitle
\section{Introduction}

\begin{figure*}
   \centering{
   \includegraphics[height=6cm]{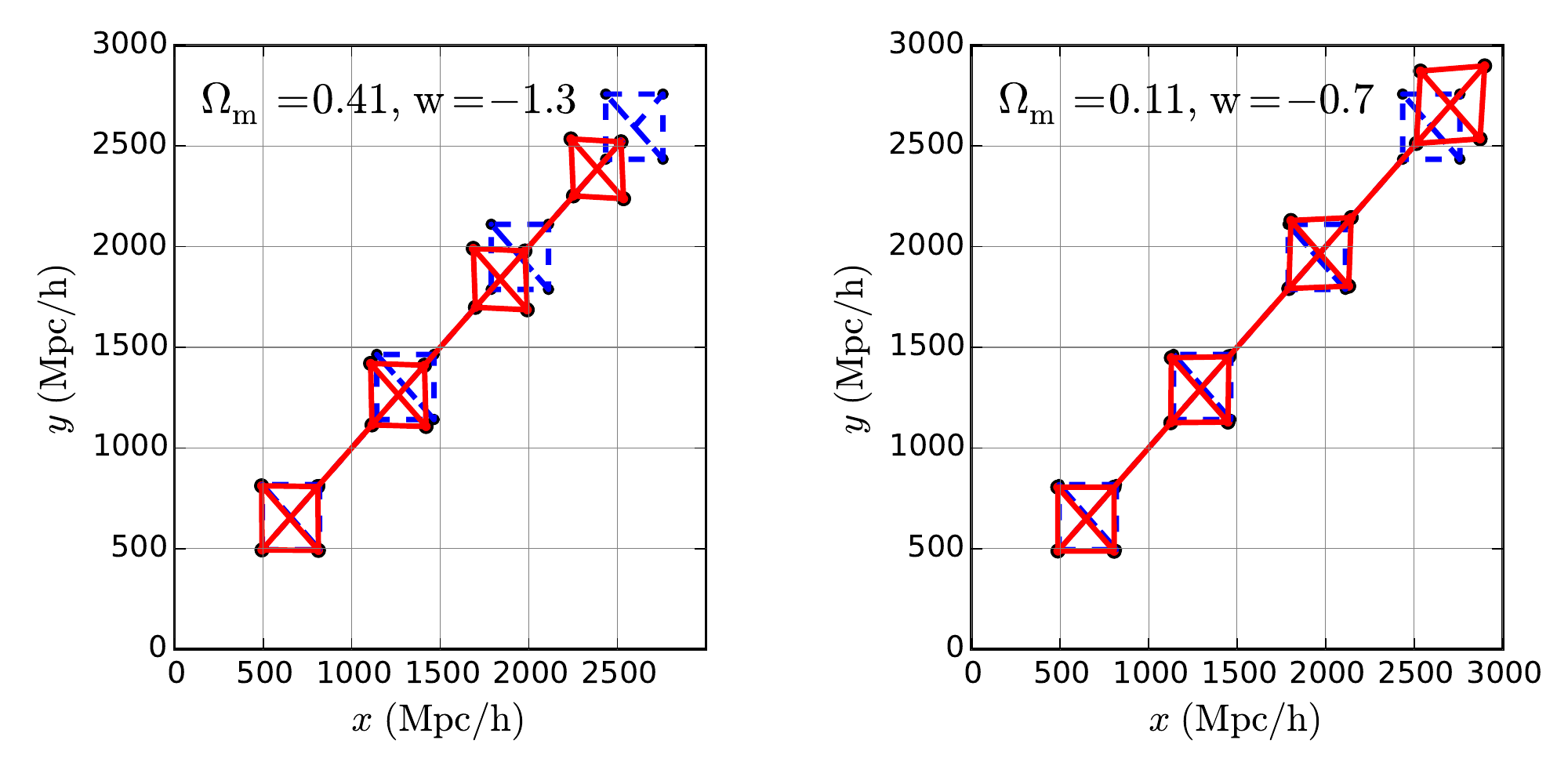}
   }
   \caption{\label{fig1}
   The AP distortion in two wrongly assumed cosmologies $\Omega_m=0.41$, $w=-1.3$ and $\Omega_m=0.11$, $w=-0.7$,
   for an observer whose true cosmology being $\Omega_m=0.26$, $w=-1$.
   Four perfect squares are measured by the observer located at the origin.
   The apparently distorted shapes are plotted in red solid lines, while the underlying true shapes are plotted in blue dashed lines.
   The large redshift dependence of the distortion motivates us to conduct the {\it tomographic} analysis.
   }
\end{figure*}

\begin{figure*}
   \centering{
   \includegraphics[width=16cm]{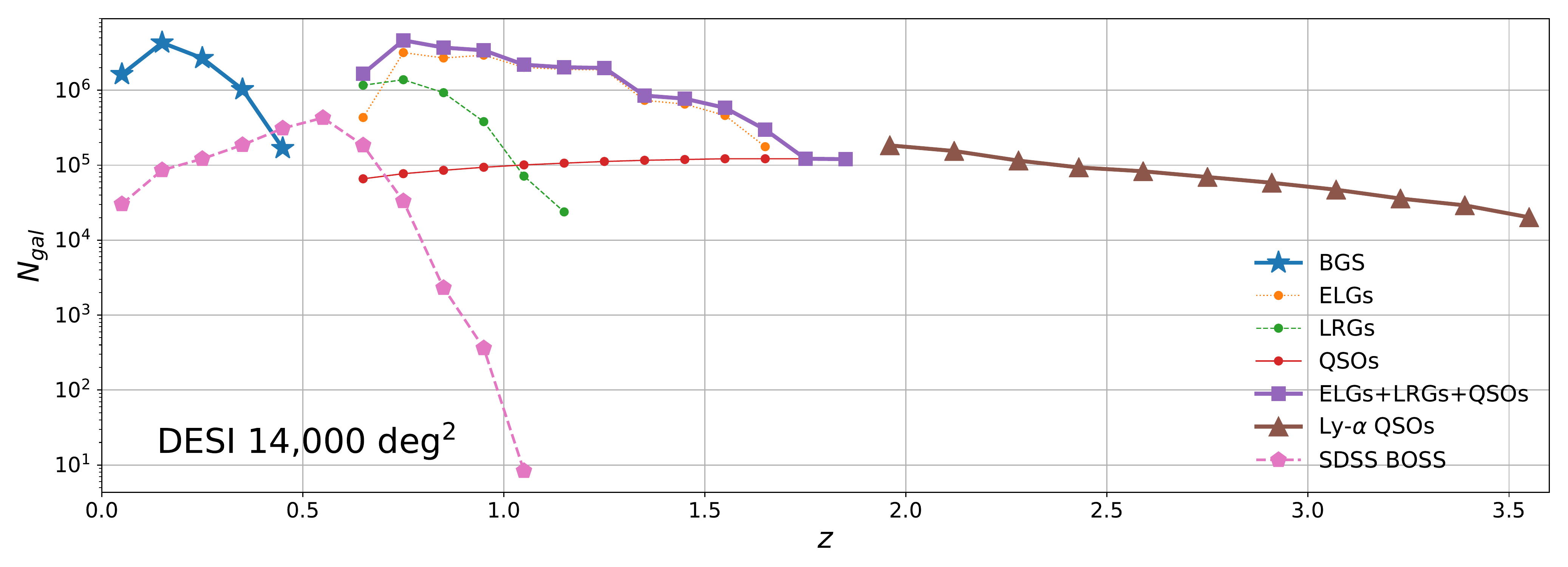}
    }
   \caption{\label{figngal} Expected redshift distribution of the galaxy sample from the DESI survey.
   We plotted the number of galaxies $N_{gal}$ in 30 redshift bins.
   Among them, the 18 redshift bins of galaxies at $z\lesssim1.9$ are used to forecast the performance of AP if applied to them.
   For comparison, the SDSS BOSS galaxies $N_{gal}$ are also plotted.
   }
\end{figure*}

The discovery of cosmic acceleration \citep{Riess1998,Perl1999} implies
either the existence of a ``dark energy'' component in our Universe
or the breakdown of general relativity on cosmological scales \citep[see ][for a recent review]{2012IJMPD..2130002Y}.
The theoretical explanation and observational probes of cosmic acceleration have attracted tremendous attention,
and are still far from being well understood or accurately measured \citep{SW1989,Li2011,DHW2013,Miao2018}.

In an effort to probe the cosmic expansion history, large scale structure (LSS) surveys are utilized to
extract information about two key geometrical quantities;
the angular diameter distance $D_A$ and the Hubble factor $H$.
If they were precisely measured as functions of redshift,
then tight constraints can be placed on cosmological parameters,
like the matter density $\Omega_m$ and the equation of state (EoS) of dark energy $w$.

The Alcock-Paczynski (AP) test \citep{AP1979} provides a geometric probe of $D_A$ and $H$.
Given a certain cosmological model, the radial and tangential sizes of distance
objects or structures can be computed as
$\Delta r_{\parallel} = \frac{c}{H(z)}\Delta z$ and $\Delta r_{\bot}=(1+z)D_A(z)\Delta \theta$,
where $\Delta z$, $\Delta \theta$ are the observed redshift span and angular size, respectively.
When incorrect cosmological models are assumed for transforming galaxy redshifts into comoving distances,
the wrongly estimated $\Delta r_{\parallel}$ and $\Delta r_{\bot}$ induces geometric distortion (see Figure \ref{fig1}).
In galaxy redshift surveys, measuring the galaxy clustering in the radial and transverse directions
enables us to probe the AP distortion, and thus place constraints on cosmological parameters
\citep{Ryden1995,Ballinger1996,Matsubara1996,Outram2004,Blake2011,LavausWandelt2012,Alam2016,Qingqing2016}.

The main difficulty of AP test is that
the radial distances of galaxies, inferred from their observed redshifts,
are inevitably affected by the galaxy peculiar motions.
This leads to apparent anisotropies in the clustering signal even if the adopted cosmology is correct.
This effect, known as redshift-space distortions (RSD),
is usually much more significant than the AP distortion,
and is notoriously difficult to be accurately modeled in the statistics of galaxy clustering \citep{Ballinger1996,Jennings2011}.

As a complementary method to apply the AP test,
\cite{Marinoni2010} proposed to statistically study a large number of galaxy pairs and search for the deviation from a symmetric distribution of direction; however, since the peculiar velocity distorts the observed redshifts and changes the apparent tilt angles of galaxy pairs,
this method is also seriously limited by RSD \citep{Jennings2011}.
In an effort to minimize RSD contamination, the shape of void regions \citep{Ryden1995,LavausWandelt2012}  has also been
proposed as an AP probe. This approach has the advantage that the void regions are easier to model compared with dense regions,
but has limitations in that it utilizes only low density regions of the LSS, and
requires large samples to attain statistical significances and achieve competitive constraints \citep{Qingqing2016}.


Recently, a novel method of applying the AP test by investigating the {\it redshift dependence} of the distortion was proposed by \cite{Li2014,Li2015}.
The method is motivated by \cite{topology}, where the authors found adopting wrong set of cosmological parameters would produce redshift-dependent distortion in the LSS.
\cite{Li2014,Li2015} applied this idea to the AP test analysis.
The authors found that, on one hand, the anisotropies produced by the RSD effect are, although large,
maintaining a nearly uniform magnitude over a large redshift range; on the other hand,
the degree of anisotropies produced by  AP varies much more significantly.
So they developed a method searching for the AP distortion from the redshift evolution of the anisotropies in LSS.

A consequence of reducing the RSD effect is that, by avoiding the complex modeling of galaxy position and velocity distributions,
it becomes possible to use galaxy clustering on scales as small as 6-40 $h^{-1}$Mpc.
In this region, there exists a large amount of clustered structures \citep{Zhang2018};
thus enables us to derive tight constraint on cosmological parameters.
This large amount of information can hardly be utilized by other LSS statistical methods.

The first application of this AP method (hereafter the tomographic AP method) to observational data was performed in \cite{Li2016}.
The authors split the 1.13 million Sloan Digital Sky Survey (SDSS) Data Release 12 (DR12) galaxies into six redshift bins,
measured their anisotropic 2PCFs, and quantified the redshift evolution of anisotropy.
In combination with the datasets of Cosmic Microwave Background (CMB), type Ia supernovae (SNIa), baryon acoustic oscillations (BAO), and the  $H_0$ measurements, the authors obtained $\Omega_m = 0.301 \pm 0.006,\ w=-1.054 \pm 0.025$
in a flat universe with cold dark matter and constant EoS dark energy components (hereafter $w$CDM).
The error bars are reduced by as much as 40\% by adding the AP method into the combination of CMB+SNIa+BAO+$H_0$.

As a follow-up study, \cite{Li2018} improved the method by developing a technique accurately approximating the 2PCFs in different cosmologies.
This greatly reduces the computational expense of the 2PCFs,
thus enables the exploration of models with three or more parameters.
\cite{Li2018} applied the method to constrain a model of dynamical dark energy EoS
$w(z)=w_0+w_a \frac{z}{1+z}$ (hereafter $w_0 w_a$CDM),
and improved the Planck+BAO+SNIa+$H_0$ constraint on $w_0$-$w_a$ by as much as 50\%.
Furthermore, in a very recent work, \cite{Zhang2018} combined the tomographic AP method with the BAO measurements,
and obtained a Hubble constant $H_0=67.78^{+1.21}_{-1.86}$ km s$^{-1}$ Mpc$^{-1}$ (2.26\% precision).
The inclusion of AP reduces the error bar by 32\%.

As a newly developed technique, the tomographic AP method shows promising potential in constraining cosmological parameters.
However, it is still far from becoming as mature as the BAO method.
To summarize, the method needs to be improved in three aspects:
\begin{itemize}
 \item We shall improve its methodology, by enhancing its statistical power,
 better understanding and estimating the systematical effects, and reducing the computational cost, etc.
 \item So far the method has only been used to constrain a limited set of parameters of $\Omega_m$, $w(z)$ and $H_0$.
 It is undoubtedly desirable to extend its application to more parameters and models.
 \item Given that there are many on-going and planned LSS surveys including DESI \citep{DESI}, EUCLID \citep{EUCLID}, HETDEX \citep{HIL}),
 it is necessary to forecast the constraining power when the method is applied to these surveys.
 \end{itemize}

In this work we explored all these three issues.
In Sec. \ref{sec:methodology},
we showed that the methodology used in \cite{Li2014, Li2015, Li2016, Li2018} neglected the correlations among different redshift bins,
which leads to an over-estimation of statistical power and a large statistical fluctuation.
We proposed a full covariance matrix approach to solve this problem and make the analysis statistically more perfect.
In Sec. \ref{sec:current}, we performed a comprehensive study on its constraints on a serious of cosmological parameters.
In Sec. \ref{sec:FutureSurvey}, we forecast the cosmological constraints expected from the DESI survey.
Conclusions of our work are given in Sec. \ref{sec:conclusion}.


\section{Data and Methodology }
\label{sec:methodology}

The data and methodology used in this work is similar to what used in \cite{Li2014,Li2015,Li2016,Li2018},
except that we more completely evaluate the statistical uncertainties.

\subsection{Data}

\subsubsection{SDSS DR12 Galaxies}
The Sloan Digital Sky Survey \citep{York2000}, as the currently largest spectroscopic galaxy survey, has obtained spectra for more than three million astronomical objects.
This created the most detailed three-dimensional maps of the Universe ever made.
BOSS (Baryon Oscillation Spectroscopic Survey) \citep{Dawson et al. 2012,Smee2013},
as a part of the SDSS-III survey \citep{Eisenstein et al. 2011},
has obtained spectra and redshifts
of 1.37 million galaxies selected from the SDSS imaging,
covering a sky region of 9\,376 $\rm deg^2$ and a redshift span of $0.1\lesssim z\lesssim0.75$.
Its wide redshift coverage and large amount of galaxies makes it the best material for performing the tomographic  analysis.

Following \cite{Li2016}, we use the spectroscopic galaxy sample of SDSS-III BOSS DR12,
containing the LOWZ catalogue at $0.1\lesssim z\lesssim0.45$ and the CMASS catalogue covering $0.4\lesssim z\lesssim0.7$ \citep{Reidetal:2016}.
For purpose of a tomographic clustering analysis, we split the sample into six, non-overlapping redshift bins of
$0.150<z_1<0.274<z_2<0.351<z_3<0.430<z_4<0.511<z_5<0.572<z_6<0.693$
\footnote{The boundaries are determined so that the number of galaxies are roughly the same in different redshift bins
(for LOWZ and CMASS samples, respectively).}.
The total number of galaxies used in the analysis is 1,133,326.

\subsubsection{Horizon Run 4 mocks}

We rely on the Horizon Run 4 \citep[HR4;][]{HR4} to estimate and correct the systematics. HR4 is a large N-body simulation with box size $L={3150}$ $h^{-1}$Mpc and number of particles $6300^3$, produced under the WMAP5\citep{komatsu2011} cosmological parameters
$(\Omega_{b},\Omega_{m},\Omega_\Lambda,h,\sigma_8,n_s)$  = (0.044, 0.26, 0.74, 0.72, 0.79, 0.96).
Mock galaxy samples are then created using a modified version of the one-to-one correspondence scheme \citep{hong2016}.
Comparing the 2pCF of the mocks to the SDSS DR7 volume-limited galaxy sample \citep{zehavi2011},
we found the simulated 2pCF shows a finger of god (FOG) feature \citep{FOG} rather close to the observation.
The projected 2pCF agrees with the observation within 1$\sigma$ deviation on scales greater than 1 ${h^{-1}}$Mpc \cite{hong2016}.

\subsubsection{MultiDark PATCHY mocks}

We utilize the set of 2,000 MultiDark PATCHY mock catalogues \citep{MDPATCHY} from the dark matter simulation to the covariance matrix.
The MultiDark PATCHY mocks are produced using approximate gravity solvers and analytical-statistical biasing models,
calibrated to the BigMultiDark N-body simulation \citep{K2014}.
The mock surveys can well reproduce the number density,
selection function, survey geometry, and 2PCF measurement of the BOSS DR12 catalogues,
and have been adopted in a series of works \citep[see][and references therein]{Alam2016} to conduct clustering analysis of BOSS galaxies.

\subsubsection{DESI galaxies}

The DESI \cite{DESI} observational program is a future project measuring the baryon acoustic feature of the large-scale structure,
as well as the distortions effects of redshift space.
DESI provides high precision measurements of the Universe's expansion rate up to z $\sim$ 1.5.
The baseline assumption is that it runs over an approximately five years period covering 14,000 deg$^2$ in area.
The DESI survey makes observations of four types of objects:
\begin{itemize}
 \item A magnitude-limited Bright Galaxy Survey (BGS)($0.05<z<0.45$) comprising approximately 10 million galaxies;
 \item Bright emission line galaxies (ELGs) (up to $z=1.65$) probing the Universe out to even higher redshift;
 \item Luminous red galaxies (LRGs) (up to $z=1.15$), which extend the BOSS LRG survey in both redshift and survey area;
 \item Quasi-stellar objects (QSOs) as direct tracers of dark matter in the redshift range $0.65<z<1.85$.
\end{itemize}
The number density distribution of these galaxies is shown in Figure \ref{figngal}. Our forecast in section 4. is based on these numbers.


\subsection{Methodology}
\label{sec:APeffect}

\subsubsection{Quantifying the Anisotropy}

\cite{Li2016} (hereafter Li16) split the BOSS DR12 galaxies into six redshift bins,
and computed the integrated 2pCF in each bin
\begin{equation}
 \xi_{\Delta s} (\mu) \equiv \int_{s_{\rm min}}^{s_{\rm max}} \xi (s,\mu)\ ds,
\end{equation}
where the correlation function $\xi$ is measured as a function of $s$, the distance separation of the galaxy pair,
and $\mu=\cos(\theta)$, with $\theta$ being the angle between the line joining the pair of galaxies and
the line of sight (LOS) direction to the target galaxy.
The range of integration was chosen as $s_{\rm min}=6$ $h^{-1}$Mpc and $s_{\rm max}=40$ $h^{-1}$Mpc.
By focusing on the redshift dependence of anisotropy, the RSD effect is largely reduced,
and it becomes possible to use the galaxy clustering down to 6 $h^{-1}$Mpc.

To mitigate the systematic uncertainty from {\it galaxy bias} and {\it clustering strength},
Li16 further normalized the amplitude of $\xi_{\Delta s}(\mu)$ to focus on the shape, i.e.
\begin{equation}\label{eq:norm}
 \hat\xi_{\Delta s}(\mu) \equiv \frac{\xi_{\Delta s}(\mu)}{\int_{0}^{\mu_{\rm max}}\xi_{\Delta s}(\mu)\ d\mu}.
\end{equation}
A cut $\mu<\mu_{\rm max}$ is imposed to reduce the fiber collision and FOG effects.

\subsubsection{The Redshift Evolution of Anisotropy}

The redshift evolution of anisotropy, between the $i$th and $j$th redshift bins are be quantified as
\begin{equation} \label{eq:deltahatxi}
\delta \hat\xi_{\Delta s}(z_i,z_j,\mu)\ \equiv\ \hat\xi_{\Delta s}(z_i,\mu) - \hat\xi_{\Delta s}(z_j,\mu),
\end{equation}
where the systematics of $\delta \hat\xi_{\Delta s}$ (hereafter $\delta\hat\xi_{\Delta s, \rm sys}$),
mainly comes from the redshift evolution of RSD effect
\footnote{The redshift evolution of anisotropy from RSD is, in general, smaller than those from AP (\cite{Li2014,Li2015}).
But it still creates small redshift dependence in $\hat \xi_{\Delta s}(\mu)$ \citep{Li2014,Li2015}.
},
was measured from the Horizon Run 4 \citep{HR4} N-body simulation and subtracted.

\subsubsection{``Part-cov'' approach of likelihood}

To quantify the overall  redshift evolution in the sample,
Li16 chose the {\it first redshift bin} as the reference, compare the measurements in higher redshift bins with it, and then sum up these differences.
So we have the following $\chi^2$ function describing the total of evolution
\begin{equation}\label{eq:chisq1}
\chi_{\rm AP,part-cov}^2\equiv \sum_{i=2}^{6} \sum_{j_1=1}^{n_{\mu}} \sum_{j_2=1}^{n_{\mu}} {\bf p}(z_i,\mu_{j_1}) ({\bf Cov}_{i}^{-1})_{j_1,j_2}  {\bf p}(z_i,\mu_{j_2}),
\end{equation}
where $n_{\mu}$ denotes the binning number of $\hat\xi_{\Delta s}(\mu)$,
and ${\bf p}(z_i,\mu_{j})$ is defined as
\begin{eqnarray}\label{eq:bfp}
 {\bf p}(z_i,\mu_{j}) \equiv&\ \delta \hat\xi_{\Delta s}(z_i,z_1,\mu_j) - \delta \hat\xi_{\Delta s, \rm sys}(z_i,z_1,\mu_j).
\end{eqnarray}
The covariance matrix ${\bf Cov}_i$ is estimated from the 2,000 MultiDark-Patchy mocks \citep{MDPATCHY}.
In wrong cosmologies, the AP effect produces large evolution of clustering anisotropy,
thus would be disfavored due to a large $\chi^2$ value.

In Eq.\ref{eq:chisq1} we labeled the $\chi^2$ function by the $\chi^2$ analysis by ``part-cov'',
to denote that this $\chi^2$ analysis
does not include the correlations among different ${\bf p}(z_i)$s.
The different ${\bf p}(z_i)$s are actually correlated with other,
in the sense that
1) since all the ${\bf p}(z_i)$s takes the first redshift bin as the reference,
the fluctuation in the first bin enters all ${\bf p}(z_i)$s and makes them statistically correlate with each other;
2) the LSS at different redshift bins have correlations even if they are not overlapping with each other
(galaxies lying near the boundary of a redshift bin have been affected by galaxies in the both nearby bins
in the past structure formation era).

We tested and found ignoring 2) does not lead to significant changes in the results, so it is a minor effect.
But ignoring 1) leads to $\sim$20\% mis-estimation of the statistical error.
The ``part-cov'' method relies on the first redshift bin as the reference.
If this bin happened to have a large deviation (i.e. due to statistical fluctuation) from its statistical expectation,
then all ${\bf p}$s would be affected.
This creates statistical error in the results, which haven't been included in \cite{Li2016,Li2018}.

Appendix A shows the mis-estimation and large fluctuation of this part-cov approach.
They would be overcome if we include the correlations among the ${\bf p}(z_i)$s into the analysis.

\subsubsection{``Full-cov'' approach of likelihood}\label{sec:fullcov}

We adopt the following formula to represent the $\chi^2$ function including all correlations,
\begin{equation}\label{eq:chisq2}
\chi_{\rm AP,full-cov}^2\equiv {\bf P} \cdot {\bf Cov} \cdot {\bf P},
\end{equation}
where
\begin{equation}\label{eq:chisq2_P}
{\bf P} = ({\bf \tilde{p}}(z_2,\mu_{j_1}),..., {\bf \tilde{p}}(z_6,\mu_{j_1})),
\end{equation}
a vector containing $(n_z-1)\times(n_{\mu})$ components,
is built by joining all ${\bf \tilde{p}}(z_i,\mu_{j})$s together.
Here we re-define the $\bf p$ as the evolution between the {\it nearby redshift bins}, i.e.
\begin{equation}\label{eq:chisq2_p}
 \tilde{\bf p} \equiv\ \delta \hat\xi_{\Delta s}(z_i,z_{i-1}) - \delta \hat\xi_{\Delta s, \rm sys}(z_i,z_{i-1}).
\end{equation}
Actually, the results does not change if we still use the original $\bf p$ defined in Eq(5)
\footnote{Once we include the full covariance matrix, the result does change
no matter how we {\it linearly transform} the $\bf p$s.}.
We redefine $\bf p$ as $\bf \tilde{p}$ so that the formula explicitly has no special redshift bin chosen as the reference.

\begin{figure}
   \centering{
   \includegraphics[width=8cm ]{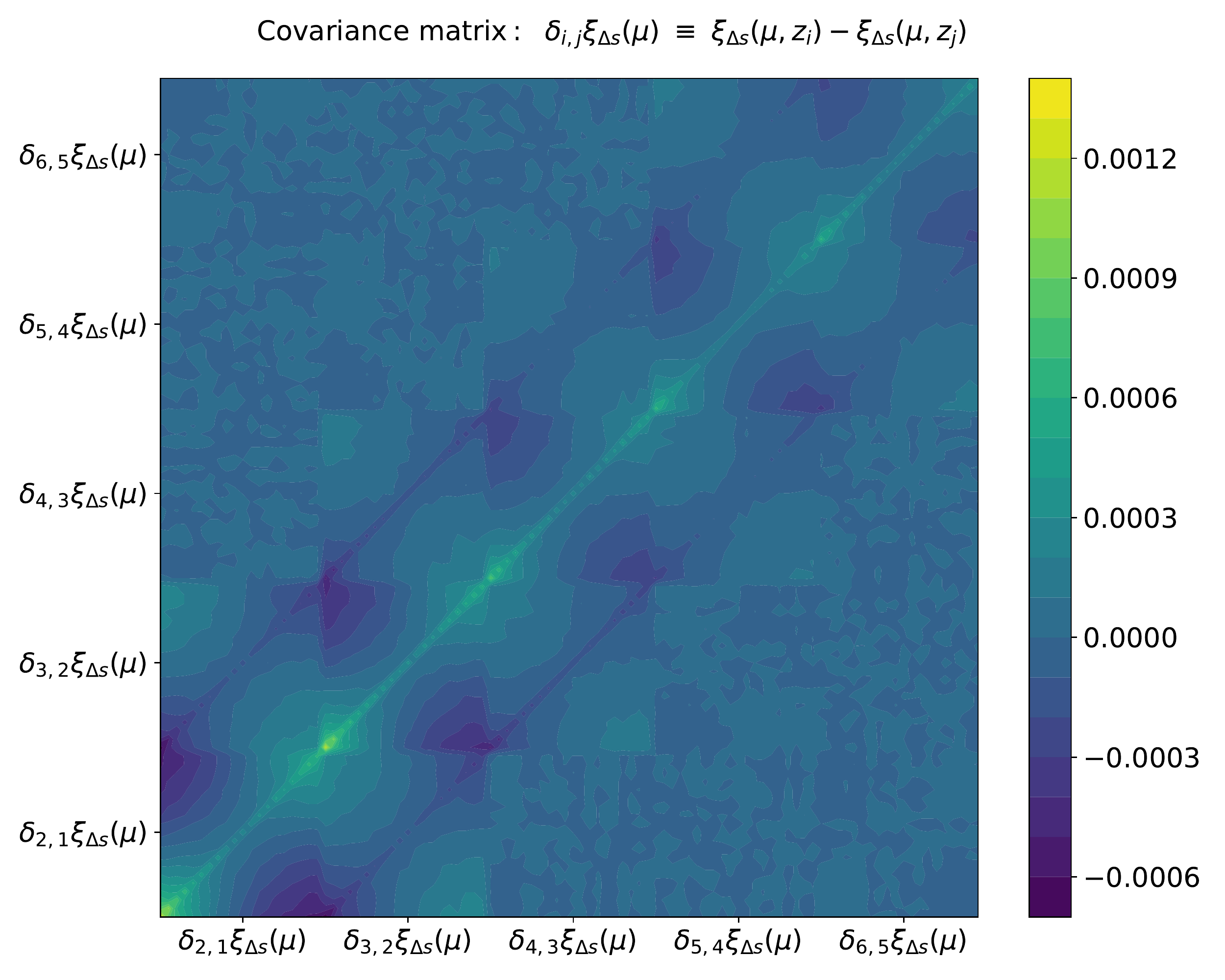}
   \includegraphics[width=8cm ]{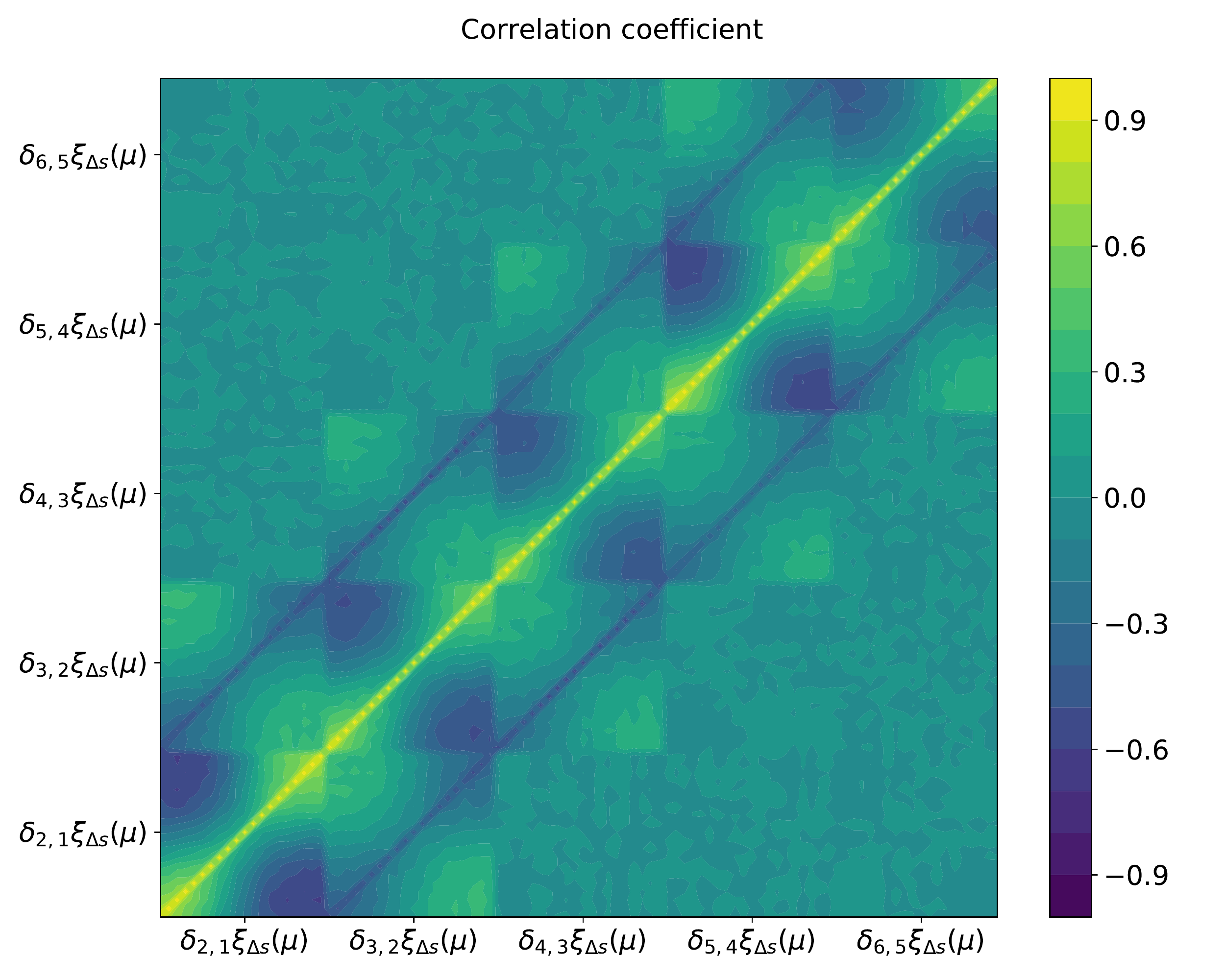}
   }
   \caption{\label{figcovmat}
   The covariance matrix of the $\hat\xi_{\Delta s}(z_i,z_{i-1})$s (upper panel),
   and its normalized version (i.e. correlation coefficients, lower panel)
   We split each $\hat\xi_{\Delta s}(z_i,z_{i-1})$ into 20 bins in the $\mu$ space,
   so five of them form a 100$\times$100 covariance matrix.
   The five 20$\times$20 matrices at the diagonal are the auto-correlations,
   while the off-diagonal matrices describe the cross-correlation.
   The non-zero cross-correlation, especially between the $\hat\xi_{\Delta s}(z_i,z_{i-1})$s next to each other,
   suggests the necessity of including them into the analysis.
   See Section \ref{sec:fullcov} for details.
   }
\end{figure}

\subsubsection{The covariance matrix}

The covariance matrix $\bf Cov$, estimated from the MultiDark mocks,
was shown in the Figure \ref{figcovmat}.
The upper panel shows the covariance matrix
when we split $\xi_{\Delta s}$ into 20 bins in $\mu$ space,
while the lower panel shows the normalized covariance matrix
(i.e. the correlation coefficient).

We find:
\begin{itemize}
 \item  Since we have 6 redshift bins, in total we need 6-1=5
 $\delta_{i,j}\xi_{\Delta s}$s to characterize the evolution among them.
  So the plot of total covariance matrix contains 5$\times$5=25
  regions of ``cells''.
 \item The five ``diagonal cells''
  describe the $20\times20$ auto covariance matrix of
  $\delta_{i,j}\xi_{\Delta s}(\mu)$.
  $\mu$-bins close to each other strong positive correlations,
  while $\mu$-bins very far away from each other
  have negative correlations imposed by the normalization condition.
 \item The 20 ``non-diagonal boxes'' describe the cross-correlation
  among different $\delta_{i,j}\xi_{\Delta s}$s.
  They have non-zero values.
 \item Those $\delta_{i,j}\xi_{\Delta s}$s who have overlapping redshift bins are strongly correlated
  since they contain the same $\xi_{\Delta s}(\mu,z_i)$
  (e.g., $\delta_{2,1}\xi_{\Delta s}$ and $\delta_{3,2}\xi_{\Delta s}$
  both depend on the $\xi_{\Delta s}$).
  Their pattern of correlation is similar to the diagonal boxes,
  i.e. positive correlation among nearby $\mu$-bins and negative among
  very faraway $\mu$-bins.
 \item  $\delta_{i,j}\xi_{\Delta s}$s without any overlapping bins do not show significant cross-correlations (correlation coefficients close to 0).
  We tested and found that, ignoring them does not have statistically significant impact on the derived cosmological constraints.
\end{itemize}

\section{Cosmological Constraints}
\label{sec:current}

The cosmological constraints are
derived from the likelihood method described in Sec. \ref{sec:APeffect}.
We divide our discussion into two sections.
In Sec. \ref{sec:de} we presented the constraints on the {\it background parameters}
of $w$, $w_0$, $w_a$, $\Omega_m$ and $H_0$,
within the framework of $\Lambda$CDM, $w$CDM and $w_0 w_a$CDM models, respectively.
These parameters are {\it directly} constrained by the AP method,
which measures the geometry of the universe in the late-time expansion era
\footnote{There is one exception. AP method alone can not put constraint on $H_0$.
The change in $H_0$ corresponds to a uniform re-scaling of LSS, and produces no anisotropy.
But the AP method can improve its constraint by breaking its degeneracy with other parameters \citep{Zhang2018}. }.
In Sec. \ref{sec:exts}, we extend the scope and explore the {\it other cosmological parameters},
including the $\Lambda$CDM parameters  and their derivations,
and the 1-parameter extensions to $\Lambda$CDM and $w$CDM models.

We present the cosmological constraints when the AP likelihood is combined with several external datasets,
including the full-mission Planck observations of CMB temperature and polarization anisotropies \citep{Planck2015},
the BAO distance priors measured from SDSS DR11 \citep{Anderson2013}, 6dFGS \citep{6dFGS} and SDSS MGS \citep{MGS},
the ``JLA'' SNIa sample \citep{JLA},
and the Hubble Space Telescope measurement of $H_0=70.6\pm3.3$ km/s/Mpc \citep{Riess2011,E14H0}.
They are exactly the same datasets used in \cite{Li2016,Li2018}.

\begin{table*}
\caption{Mean values and 68\% confidence limits of cosmological parameters for the
$\Lambda$CDM, $w$CDM and $w_0 w_a$CDM models,
from combinations of Planck+BAO and Planck+BAO+AP.
The uncertainties (of the last 2 digits of the numbers) are listed in the brackets.}
\label{Table:DE}
\begingroup
\openup 5pt
\newdimen\tblskip \tblskip=5pt
\nointerlineskip
\vskip -3mm
\scriptsize
\setbox\tablebox=\vbox{
    \newdimen\digitwidth
    \setbox0=\hbox{\rm 0}
    \digitwidth=\wd0
    \catcode`"=\active
    \def"{\kern\digitwidth}
    \newdimen\signwidth
    \setbox0=\hbox{+}
    \signwidth=\wd0
    \catcode`!=\active
    \def!{\kern\signwidth}
\halign{ \hbox to 1.3in{$#$\leaderfil}
\tabskip=0.1em &\hfil $#$ \hfil & \hfil $#$ \hfil & \hfil $#$ \hfil & \hfil $#$ \hfil & \hfil $#$ \hfil  & \hfil $#$ \hfil & \hfil $#$ \hfil & \hfil $#$ \hfil & \hfil $#$ \hfil & \hfil $#$ \hfil & \hfil $#$
\cr \noalign{\doubleline}
\multispan1\hfil \hfil &
\multispan2\hfil $\Omega_{\rm m}$ \hfil &
\multispan2\hfil $H_0$ \hfil&
\multispan2\hfil $w_0$ \hfil&
\multispan2\hfil $w_a$ \hfil&
\cr \noalign{\vskip -3pt} \omit&\multispan2\hrulefill&\multispan2\hrulefill&\multispan2\hrulefill&\multispan2\hrulefill&\multispan2\hrulefill&
\cr \omit\hfil Model
  \hfil&\omit\hfil Planck+BAO\hfil&\omit\hfil +AP
  \hfil&\omit\hfil Planck+BAO\hfil&\omit\hfil +AP
  \hfil&\omit\hfil Planck+BAO\hfil&\omit\hfil +AP
  \hfil&\omit\hfil Planck+BAO\hfil&\omit\hfil +AP
  \hfil
\cr \noalign{\vskip 5pt\hrule\vskip 3pt}
  \Lambda {\rm CDM}
  & \ \ 0.3102(64)\ \
  & \ \ 0.3041(67) \ \
  &  \ \ 67.63(48) \ \
  &  \ \ 68.08(52) \ \
\cr
  w {\rm CDM}
  & 0.306(13)
  &  0.293(10)
  & 68.3\pm1.5
  &  69.8\pm1.2
  & -1.031(63)
  & -1.090(47)
\cr  w_0 w_a{\rm CDM}
  & 0.351(29)
  & 0.300(17)
  & 64.0\pm2.7
  & 69.3\pm1.7
  & -0.51(30)
  & -0.95(18)
  & -1.47(83)
  & -0.53(53)
\cr \noalign{\vskip 5pt\hrule\vskip 3pt}
} 
} 
\endPlancktable
\endgroup
\end{table*}

\subsection{Constraints on background parameters}
\label{sec:de}

Table \ref{Table:DE} summarizes the Planck+BAO and Planck+BAO+AP constraints on the background parameters $w$, $w_0$, $w_a$, $\Omega_m$ and $H_0$,
within the framework of $\Lambda$CDM, $w$CDM and $w_0 w_a$CDM models.
In what follows, we discuss about them in details.

\subsubsection{$\Lambda$CDM parameters}

The $\Lambda$CDM model with EoS $w_{\Lambda}=-1$ is the simplest candidate
among a large number of dark energy models, with the Hubble parameter taking form of
\begin{equation}
H(z)=H_0[\Omega_m (1+z)^3+\Omega_{\Lambda}]^{1/2},
\end{equation}
where $\Omega_m+\Omega_{\Lambda}=1$ (we neglect curvature and radiation).
Although $\Lambda$CDM model seriously suffers from the theoretical fine tuning and coincidence problems \citep{SW1989},
it is in good agreement with most of the current observational data \citep{Li2011}.

Table 1 list the constraints on $\Omega_{m}$, $H_0$ derived from Plank+BAO and Plank+BAO+AP, respectively.
Including AP into the analysis leads to a $\lesssim1\sigma$ shift in the central values of $\Omega_m$, $H_0$,
i.e. from (0.310,67.6) to (0.304,68.1), respectively.

\subsubsection{$w$CDM parameters}

\begin{figure}
   \centering{
   \includegraphics[width=7cm]{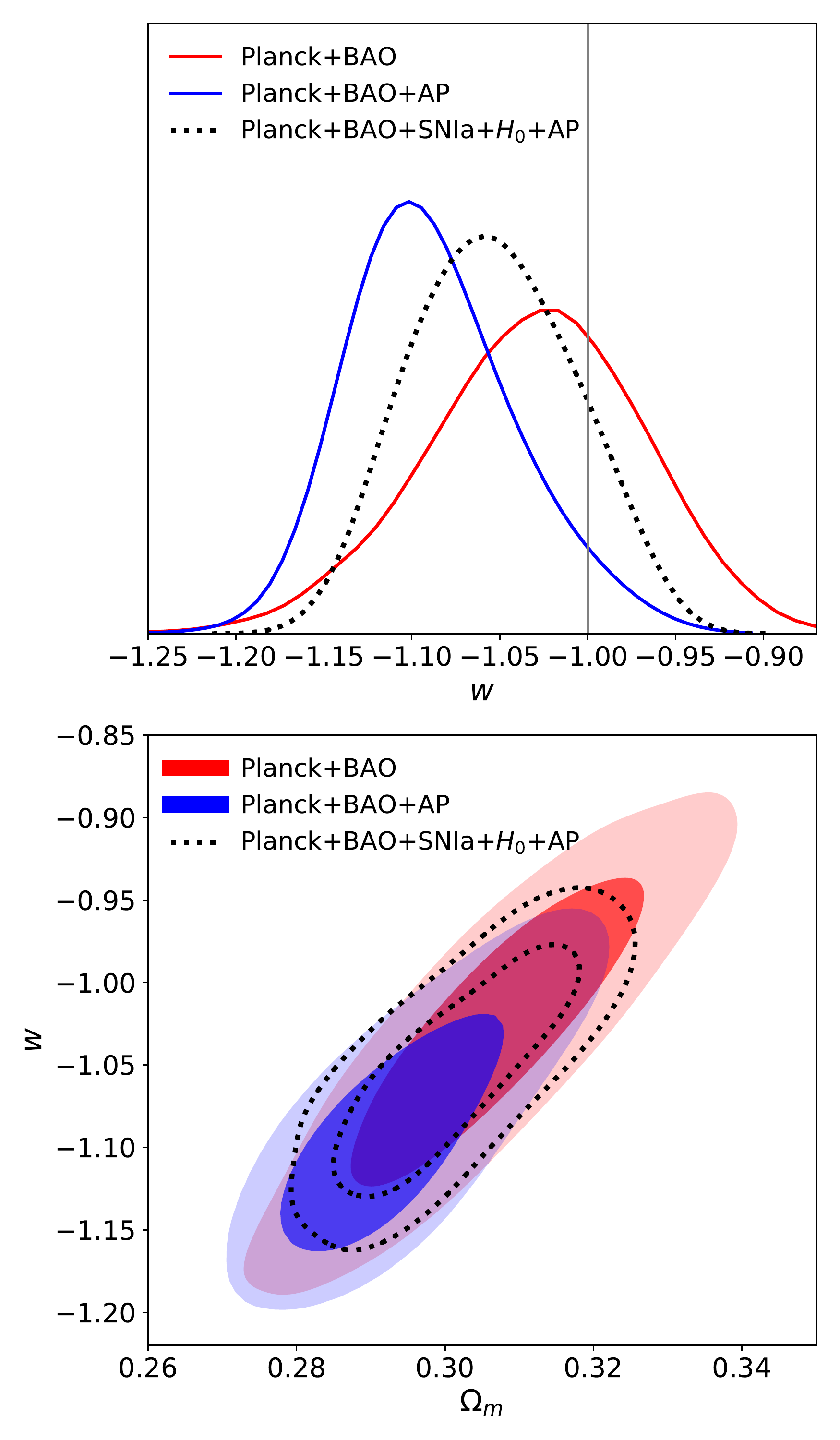}
    }
   \caption{\label{figwcdm} Cosmological constraints within the framework of $w$CDM.
        Upper panel: the constraint on $w$ from the combinations of
   	Planck+BAO, Planck+BAO+AP, Planck+BAO+SNIa+$H_0$+AP.
   	Lower panel: likelihood contours (68.3 and 95.4\% CL) in the $\Omega_m$-$w$ plane.
   }
\end{figure}

\begin{figure}
   \centering{
   \includegraphics[width=7cm]{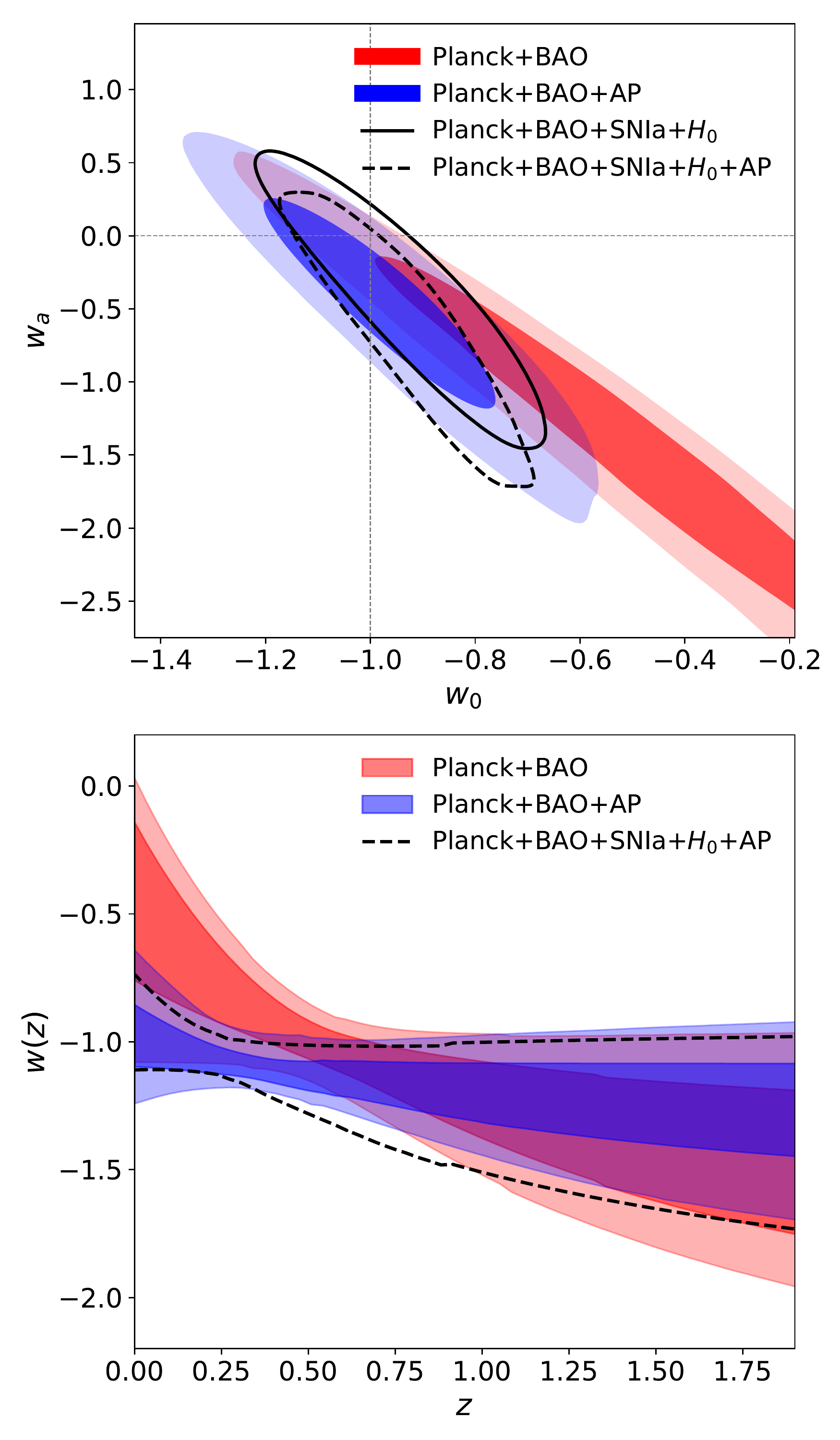}
       }
   \caption{\label{figcpl} Cosmological constraints for $w_0 w_a$CDM model.
   Upper panel: likelihood contours (68.3 and 95.4\%) in the $\Omega_m$-$w$ plane.
   Adding AP into Planck+BAO greatly improves the constraints on the $w_0-w_a$ parameters.
   Low panel: evolution of $w(z)$ in the 68.3 and 95.4\% CLs.
   A dynamical dark energy crossing $w=-1$ is mildly favored $\sim1.5\sigma$ CL.
   }
\end{figure}

The simplest generalization to $\Lambda$ is considering a constant dark energy EoS $w$,
and the Hubble parameter is given by
\begin{equation}
H(z)=H_0[\Omega_m (1+z)^3+\Omega_{de}(1+z)^{3(1+w)}]^{1/2},
\end{equation}
where $\Omega_{de}$ is the current value of the dark energy density.
If $w=-1$, then $w$CDM reduces to $\Lambda$CDM, with $\Omega_{de}=\Omega_{\Lambda}$.

The upper panel of Figure \ref{figwcdm} illustrates the
constraint on $w$ from the combination of
Planck+BAO, Planck+BAO+AP, Planck+BAO+SNIa+$H_0$+AP.
The mean values as well as the 68\% and 95\% limits are
\begin{eqnarray}
 &w&= -1.031\ ^{+0.067}_{-0.057}\  ^{+0.117}_{-0.130} \ \ {\rm Planck+BAO}; \\
 &w&= -1.089\ ^{+0.040}_{-0.054}\ ^{+0.104}_{-0.088} \ \ {\rm \ +AP}; \\
 &w&= -1.054\ ^{+0.046}_{-0.052}\  ^{+0.090}_{-0.085} \ \  {\rm \ +SNIa+}H_0{\rm +AP}.
\end{eqnarray}
If we describe the {\it decrement in the error bar} (or equivalently, improvement in the precision) by
$\frac{\sigma_{\rm w.o.\rm AP}-\sigma_{\rm with\ AP}}{\sigma_{\rm w.o.\ AP}}$,
then adding AP into the Planck+BAO combination reduces the errors by $\sim$30\%.
The inclusion of AP also shifts the constraint toward negative EoS by $\sim$1$\sigma$, making the results marginally consistent with $w=-1$ in 2$\sigma$.
Further adding the SNIa and $H_0$ ``pulled back'' it towards $w=-1$.

The lower panel of Figure 4 shows the marginalized constraint in the $\Omega_m$-$w$ plane.
We see a positive degeneracy between the two parameters,
and a shift of $w$ towards negative values.
Correspondingly, a smaller amount of $\Omega_m$ is preferred.

It is commonly believed that since the CMB data helps the BAO method to determine the absolute value of the sound horizon,
tight constraints can be achieved if the two combined.
So we will use Planck+BAO as a ``standard combination'' and check how much the constraints improve after adding AP.
In fact, because CMB and AP constrain different epochs of expansion history,
combining them can also effectively reduce the uncertainties of parameters.
This can be seen from Figure 10 of \cite{Li2016},
where we find the Planck and AP contours have orthogonal directions of degeneracy in the $\Omega_m$-$w$ space.
Actually, combining these two constrains give
\begin{equation}
\Omega_m = 0.295 \pm 0.015,\ H_0 = 69.7 \pm 1.7,\ w = -1.08 \pm 0.05,
\end{equation}
which are as tight as the Planck+BAO results.

The derived constrained from the full-covmat analysis is consistent with \cite{Li2016}
except that here we obtained a larger error bar, mainly due to the inclusion of full covariance.
A comparison between the two sets of results are presented in Appendix \ref{sec:Appendixes:wCDM}.

To ensure the robustness of the results,
we have conducted a serious checks about the systematics and the options of the AP  analysis.
We do not find any statistically significant effect on the derived results.
These tests are briefly discussed in Appendix \ref{sec:Appendixes:wCDM}.

\subsubsection{$w_0 w_a$CDM parameters}

We move on to a further generalization and consider a dynamical EoS depend on $z$.
As a simplest parameterization widely used in the literature,
one can consider the 1st order Taylor expansion of $w_{de}$ with respect to $(1-a)$, i.e.
\begin{equation}
w_{de}(z) = w_0 + w_a (1-a) = w_0 + w_a \frac{z}{1+z},
\end{equation}
which is the well-known Chevallier-Polarski-Linder (CPL) parameterization proposed by Refs. \cite{CPL_CP,L2002}.
The Hubble parameter is
\begin{equation}
H(z)=H_0[\Omega_m (1+z)^3 +\Omega_{de}(1+z)^{3(1+w_0+w_a)}e^{-3\frac{w_a z}{1+z}}]^{1/2}.
\end{equation}

Figure \ref{figcpl} shows the constraints on this dynamical dark energy model.
The Planck+BAO combination can not lead to effective constraints
on the $w_0$-$w_a$ parameters.
We only obtain two weak bounds of $w_0>-1.2,\ w_a<0.5$ (95\%),
and the upper bound of $w_0$ and lower bound of $w_a$ is left unconstrained.
Adding AP closes the constraints and yields to
\begin{eqnarray}
 &&w_0 = -0.95\ ^{+0.13}_{-0.18}\ ^{+0.35}_{-0.35}, \nonumber  \\
 &&w_a = -0.53\ ^{+0.57}_{-0.43}\ ^{+1.08}_{-1.17}\ \ {\rm Planck+BAO+AP}.
\end{eqnarray}
The result is consistent with $w(z)=-1$ in 1$\sigma$.
This manifests the power of AP method.
Combining it with BAO significantly
increases the amount of information extracted from the LSS data,
and greatly tightens the dark energy constraint.

When further considering the SNIa and $H_0$ datasets, we find
\begin{eqnarray}
&&w_0 = -0.94\ ^{+0.11}_{-0.11}\ ^{+0.22}_{-0.21},\nonumber  \\
&&w_a = -0.38\ ^{+0.45}_{-0.39}\ ^{+0.82}_{-0.87} \nonumber  \\
&& \ \ \ ({\rm Planck+BAO+SNIa+}H_0);  \\
&&w_0 = -0.93\ ^{+0.09}_{-0.10}\ ^{+0.20}_{-0.18}, \nonumber   \\
&&w_a = -0.65\ ^{+0.45}_{-0.40}\ ^{+0.83}_{-0.88} \nonumber   \\
&& \ \ \ ({\rm Planck+BAO+SNIa+}H_0{\rm +AP}).
\end{eqnarray}
The main effect of adding AP is a $\approx$0.7$\sigma$ shift of $w_a$ towards negative values
(which can be seen evidently from the upper panel of Figure \ref{figcpl}).
As a result, a dynamical dark energy (i.e., $w_a \neq -1$) is preferred at $\approx 1.5 \sigma$.
The lower panel of the Figure shows that,
the dark energy EoS is evolving from $<-1$ to $>-1$ from high redshift epoch to the present.
$w=-1$ is consistent with the constraint at $2\sigma$ CL
\footnote{In comparison,
\cite{Li2018} found adding AP into Planck+BAO+SNIa+$H_0$ leads to
a dynamical dark energy at $\approx 1\sigma $ CL
together with 50\% reduction of $w_0$-$w_a$ parameter space.}.




\subsection{Constraints on the other cosmological parameters}
\label{sec:exts}

\subsubsection{$\Lambda$CDM parameters}

\begin{table}
\caption{ Mean values and 68\% confidence limits of cosmological parameters
for the base $\Lambda$CDM model from Planck+BAO and Planck+BAO+AP combinations.
The uncertainties (of the last 2 digits of the numbers) are listed in the brackets.}
\label{Table:LCDM}
\begingroup
\openup 5pt
\newdimen\tblskip \tblskip=5pt
\nointerlineskip
\vskip -3mm
\scriptsize
\setbox\tablebox=\vbox{
    \newdimen\digitwidth
    \setbox0=\hbox{\rm 0}
    \digitwidth=\wd0
    \catcode`"=\active
    \def"{\kern\digitwidth}
    \newdimen\signwidth
    \setbox0=\hbox{+}
    \signwidth=\wd0
    \catcode`!=\active
    \def!{\kern\signwidth}
\halign{ \hbox to 1.3in{$#$\leaderfil}
\tabskip=0.1em &\hfil $#$ \hfil & \hfil $#$ \hfil & \hfil $#$ \hfil & \hfil $#$ \hfil & \hfil $#$ \hfil  & \hfil $#$ \hfil & \hfil $#$ \hfil & \hfil $#$ \hfil & \hfil $#$ \hfil & \hfil $#$ \hfil & \hfil $#$
\cr \noalign{\doubleline}
\multispan1\hfil \hfil &
\multispan2\hfil $\Lambda$CDM \hfil &
\cr \noalign{\vskip -3pt} \omit&\multispan2\hrulefill&\multispan2\hrulefill&\multispan2\hrulefill&\multispan2\hrulefill&\multispan2\hrulefill&
\cr \omit\hfil Parameter
  \hfil&\omit\hfil Planck+BAO\hfil&\omit\hfil +AP
  \hfil
\cr \noalign{\vskip 5pt\hrule\vskip 3pt}
  \Omega_b h^2
  & \ \ 0.02228(14) \ \
  & \ \ 0.02235(14) \ \
\cr 
  \Omega_c h^2
  & 0.1189(10)
  & 0.1179(12)
\cr 
  100 \theta_{\rm MC}
  & 1.04081(31)
  & 1.04093(32)
\cr 
  \tau
  & 0.081(17)
  & 0.086(17)
\cr 
  \ln (10^{10} A_s)
  & 3.094(33)
  & 3.102(33)
\cr 
  n_s
  & 0.9669(39)
  & 0.9695(39)
\cr \noalign{\vskip 5pt\hrule\vskip 3pt}
  H_0
  & 67.63(48)
  & 68.08(52)
\cr 
  \Omega_\Lambda
  & 0.6898(64)
  & 0.6959(78)
\cr 
  \Omega_m
  & 0.3102(64)
  & 0.3041(68)
\cr 
  \Omega_m h^2
  & 0.14182(98)
  & 0.14092(10)
\cr 
  \Omega_m h^3
  & 0.09591(30)
  & 0.09594(30)
\cr 
  \sigma_8
  & 0.829(13)
  & 0.829(14)
\cr 
  \sigma_8 \Omega_m^{0.5}
  & 0.4615(88)
  & 0.4572(89)
\cr 
  \sigma_8 \Omega_m^{0.25}
  & 0.618(10)
  & 0.616(11)
\cr 
  z_{\rm re}
  & 10.17^{+1.56}_{-1.38}
  & 10.56^{+1.58}_{-1.37}
\cr 
  10^9 A_s
  & 2.208(72)
  & 2.225(75)
\cr 
  10^9 A_s e^{-2\tau}
  & 1.877(11)
  & 1.873(11)
\cr 
  {\rm{Age}}/{\rm{Gyr}}
  &  13.806(22)
  &  13.790(23)
\cr 
  z_*
  & 1089.94(23)
  & 1089.76(24)
\cr 
  r_*
  &  144.79(24)
  &  144.98(26)
\cr 
  100\theta_*
  & 1.04100(30)
  & 1.04112(32)
\cr 
  z_{\rm{drag}}
  & 1059.64(30) 
  & 1059.75(30) 
\cr 
  r_{\rm{drag}}
  & 147.49(25)
  & 147.66(26)
\cr 
  k_{\rm D}
  & 0.14038(29)
  & 0.14025(30)
\cr 
  z_{\rm{eq}}
  & 3374(24)  
  & 3352(25) 
\cr 
   k_{\rm{eq}}
  & 0.010297(72) 
  & 0.010231(77)
\cr 
  100\theta_{\rm{s,eq}}
  & 0.4520(23)
  & 0.4541(25)
\cr \noalign{\vskip 5pt\hrule\vskip 3pt}
} 
} 
\endPlancktable
\endgroup
\end{table}

\begin{table*}
\caption{\label{Table:Exts} Constraints on 1-parameter extensions to the $\Lambda$CDM and $w$CDM models, for combinations of Planck+BAO and Planck+BAO+AP.
Note that we quote 95\% limits here.}
\begingroup
\openup 5pt
\newdimen\tblskip \tblskip=5pt
\nointerlineskip
\vskip -3mm
\scriptsize
\setbox\tablebox=\vbox{
    \newdimen\digitwidth
    \setbox0=\hbox{\rm 0}
    \digitwidth=\wd0
    \catcode`"=\active
    \def"{\kern\digitwidth}
    \newdimen\signwidth
    \setbox0=\hbox{+}
    \signwidth=\wd0
    \catcode`!=\active
    \def!{\kern\signwidth}
\halign{ \hbox to 1.3in{$#$\leaderfil}
\tabskip=0.1em &\hfil $#$ \hfil & \hfil $#$ \hfil & \hfil $#$ \hfil & \hfil $#$ \hfil & \hfil $#$ \hfil  & \hfil $#$ \hfil & \hfil $#$ \hfil & \hfil $#$ \hfil & \hfil $#$ \hfil & \hfil $#$ \hfil & \hfil $#$
\cr \noalign{\doubleline}
\multispan1\hfil \hfil &
\multispan2\hfil $\Lambda$CDM extension \hfil &
\multispan2\hfil $w$CDM extension \hfil&
\cr \noalign{\vskip -3pt} \omit&\multispan2\hrulefill&\multispan2\hrulefill&\multispan2\hrulefill&\multispan2\hrulefill&\multispan2\hrulefill&
\cr \omit\hfil Parameter
  \hfil&\omit\hfil Planck+BAO\hfil&\omit\hfil +AP
  \hfil&\omit\hfil Planck+BAO\hfil&\omit\hfil +AP
  \hfil
\cr \noalign{\vskip 5pt\hrule\vskip 3pt}
  \Omega_k
  & \ \ -0.0002 ^{+0.0041}_{-0.0040} \ \
  & \ \ 0.0004 ^{+0.0042}_{-0.0039} \ \
  & \ \ -0.0010^{+0.0066}_{-0.0061}\ \
  & \ \ -0.0015^{+0.0042}_{-0.0044}\ \
\cr 
  \sum m_\nu [{\rm eV}]
  & < 0.181 
  & < 0.141 
  & < 0.295 
  & < 0.243 
\cr 
  N_{\rm eff}
  & 2.97^{+0.34}_{-0.34}
  & 3.07^{+0.33}_{-0.33}
  & 2.95^{+0.38}_{-0.37}
  & 2.96^{+0.37}_{-0.35}
\cr 
  d n_s / d \ln k
  & -0.0023^{+0.0132}_{-0.0138}
  & -0.0025^{+0.0133}_{-0.0136}
  & -0.0024^{+0.0134}_{-0.0136}
  & -0.0025^{+0.0132}_{-0.0139}
\cr 
  r
  & < 0.115 
  & < 0.121 
  & < 0.113 
  & < 0.111 
\cr \noalign{\vskip 5pt\hrule\vskip 3pt}
} 
} 
\endPlancktable
\endgroup
\end{table*}

\begin{figure*}
   \centering{
   \includegraphics[width=16cm]{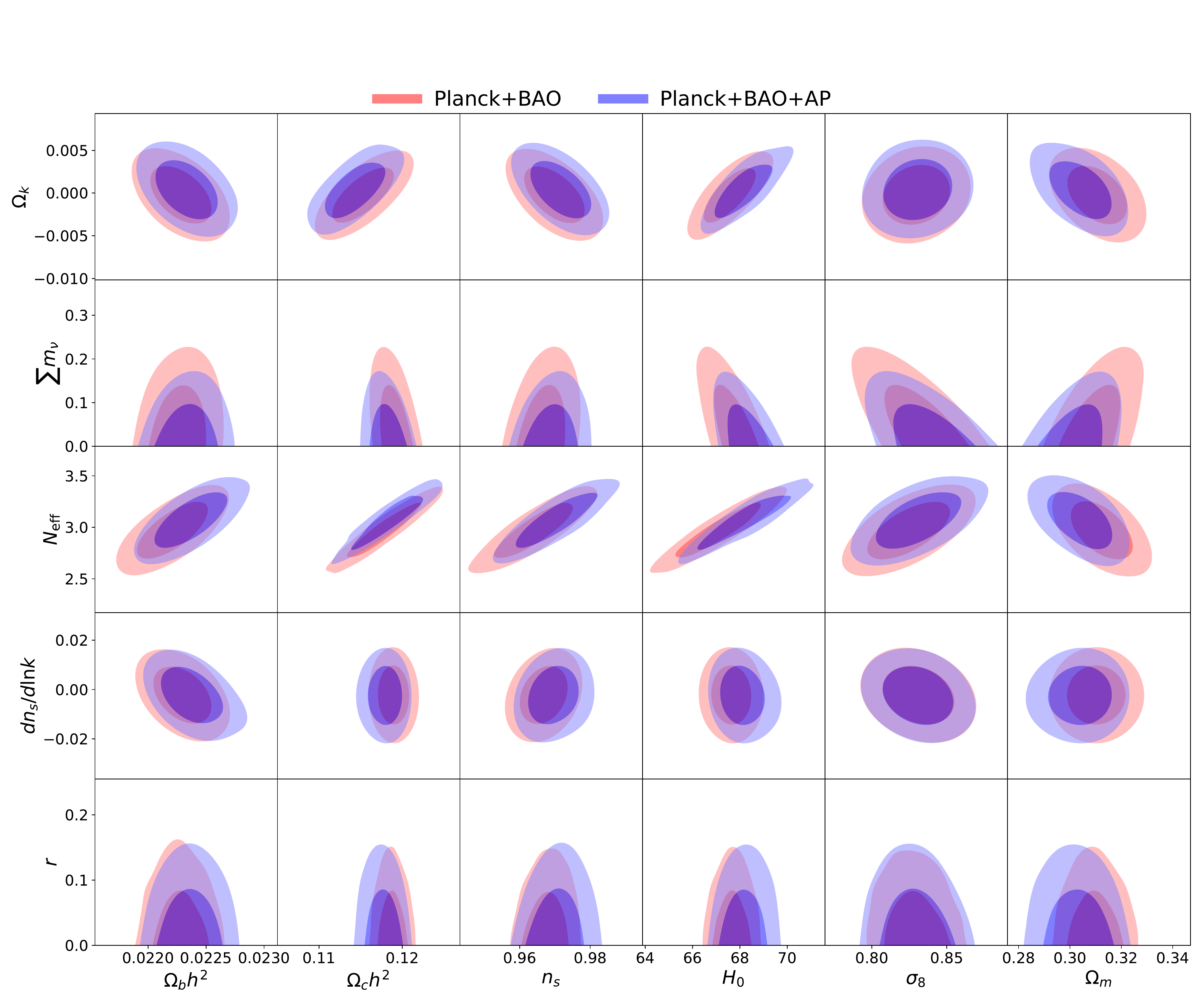}
    }
   \caption{\label{figlcdmexts} 68\% and 95\% confidence regions on
   1-parameter extensions of the base $\Lambda$CDM,
   for Planck+BAO (red) and Planck+BAO+AP (blue).
   }
\end{figure*}


\begin{figure*}
   \centering{
   \includegraphics[width=16cm]{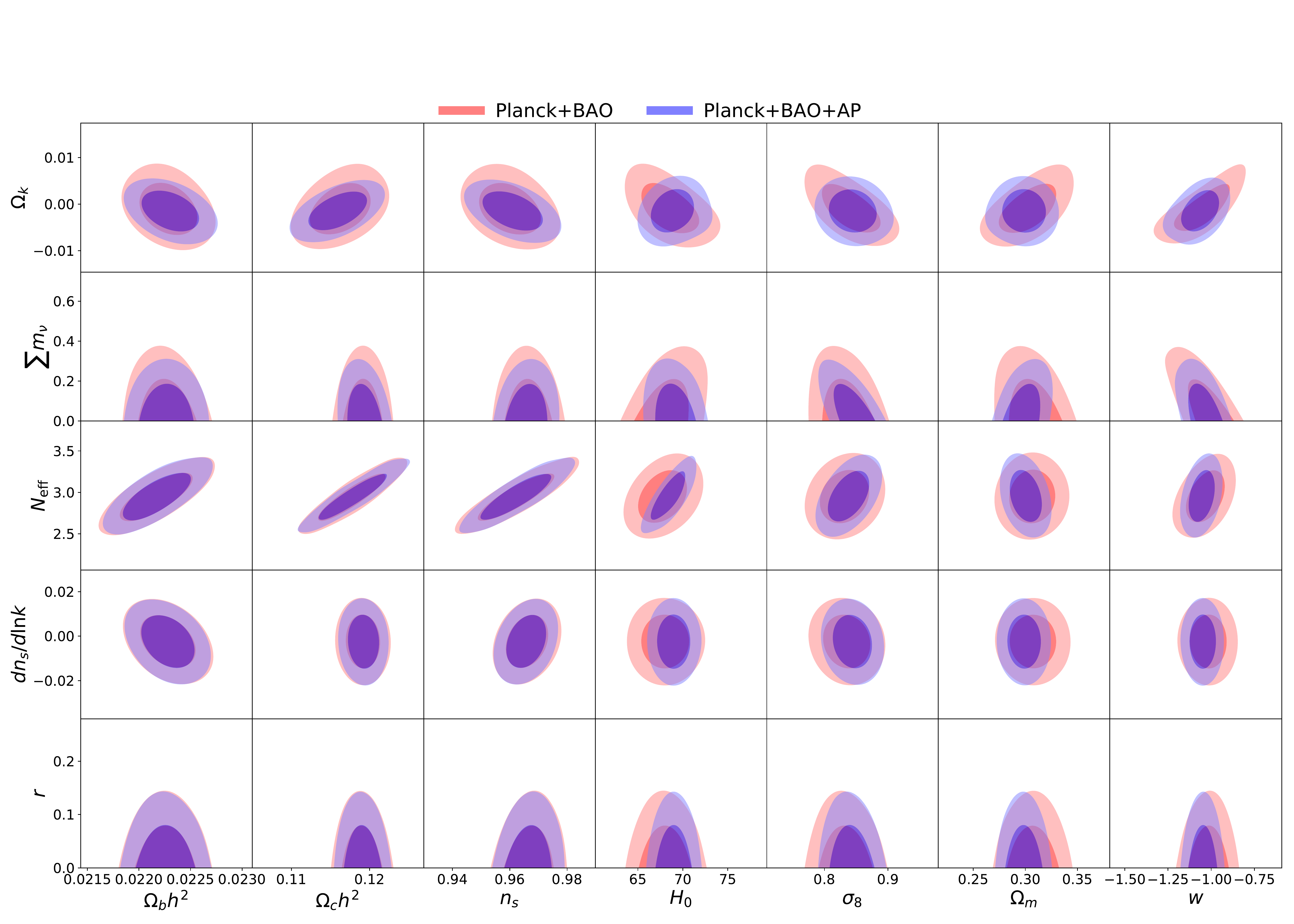}
    }
   \caption{\label{figwcdmexts} 68\% and 95\% confidence regions on
   1-parameter extensions of the $w$CDM,
   from Planck+BAO (red) and Planck+BAO+AP (blue).
   }
\end{figure*}

Table \ref{Table:LCDM} summarizes the $\Lambda$CDM parameters
(6 basic parameters, 21 derived;
see \cite{Planck2015} for the explanation of their meanings)
constrained by Planck+BAO and Planck+BAO+AP combinations.

In the $\Lambda$CDM framework, adding AP into the analysis only affects the constraint on $\Omega_m$.
The matter amount $\Omega_m$ changes from $0.310$ to 0.304,
which is a 1.0$\sigma$ CL drop
(hereafter we use the Planck+BAO error bar to quantify the CLs of the changes).
This then affects the constrains on many other parameters via the degeneracy among the parameters.
For the basic parameters, we find
\footnote{The uncertainties of some parameters become larger after AP is additionally combined.
This is due to the tension between the AP and Planck+BAO datasets.
The inclusion of AP can create a bi-peak PDF (probability density function) and the uncertainty is slightly increased
(the existence of the second peak widens the 68.3\% CL region).}:
\begin{itemize}
 \item The cold dark matter density $\Omega_c h^2$ also decreases by 1.0$\sigma$.
 \item The baryon ratio $\Omega_b h^2$ is increased by 0.5$\sigma$,
 which should come from the increasing of $H_0$.
 \item The scalar spectral index $n_s$, which has negative degeneracy
 between $\Omega_c h^2$, is decreased by 0.7$\sigma$.
 \item The Thomson scattering optical depth due to reionization, $\tau$,
 and the log power of the primordial curvature perturbations, $\ln (10^{10} A_s)$,
 has little degeneracy with the above parameters.
 So they are less affected (change < 0.3$\sigma$).
\end{itemize}
This leads to a serious of changes in the derived parameters:
\begin{itemize}
 \item Due to the negative degeneracy between $\Omega_m$ and $H_0$,
 the latter increased by 0.9$\sigma$.
 \item $\sigma_8$ is negatively correlated with both $\Omega_m$ and $H_0$;
 as a net effect, its value remains unchanged.
 \item The acoustic scale $100\theta_*$, crucially determined by the CMB angular power spectrum measurement, remains less affected.
 \item The $z_*$ and $r_*$,
 affected by the density of energy components,
 are decreased and increased by 0.8$\sigma$;
 similar effects are found for $z_{\rm drag}$ and $r_{\rm drag}$.
 \item The change in $\Omega_m$ and $H_0$ leads to corresponding change in
 the combinations of $\Omega_\Lambda$, $\Omega_m h^2$, $\sigma_8 \Omega_m^{0.5}$,
 and $\sigma_8 \Omega_m^{0.25}$.
 The only exception is $\Omega_m h^3$; its value is almost fully
 determined by the acoustic scale $\theta_*$,
 so it remains unchanged.
 \item The age of the universe is rather sensitive to $\Omega_m$ and $\Omega_\Lambda$;
 we find it decreased by $0.7\sigma$.
 \item The matter-radiation equality redshift $z_{\rm eq}$ drops by 0.9$\sigma$,
 i.e. happens in latter epoch.
 It is due to the drop in $\Omega_m$ and increment in radiation density $\Omega_r$
 (because of larger $h$).
 \item Affected by $z_{\rm eq}$ and the fraction of energy components,
 $k_{\rm eq}\equiv a(z_{\rm eq})H(z_{\rm eq})$
 and $\theta_{s,\rm eq}\equiv r_s(z_{\rm eq})/D_A(z_star)$
 (the comoving wavenumber of perturbation mode that
 entered Hubble radius at $z_{\rm eq}$,
 and the angular scale of the sound horizon at $z_{\rm eq}$)
 are also changed by 0.9$\sigma$.
 \item The characteristic wavenumber for damping $k_{\rm D}$
 (which determines the photon diffusion length),
 whose value is related to the fraction of energy components,
 is slightly changed by 0.4$\sigma$.
 \item Parameters directly determined by $\tau$ and $A_s$, including
 $10^9A_s$,  the parameter $10^9 A_s e^{-2\tau}$ describing
 small-scale damping of CMB due to Thomson scattering at reionization,
 and the reionization redshift $z_{\rm re}$, all remain less affected.
\end{itemize}





\subsubsection{1-Parameter Extensions to $\Lambda$CDM and $w$CDM}

The Planck+BAO and Planck+BAO+AP constraint on other parameters of
one-parameter extensions to $\Lambda$CDM and wCDM cosmology are given in Table \ref{Table:Exts}.
The determination of uncertainty listed in the table are at a confidence level of 95\%.

In this case, adding AP into the combination decreases the value of $\Omega_m$,
pulls the constrained region of $w$ to slightly negative values,
and also reduces its error bar by 30\%.
This leads to evident effect on the curvature $\Omega_k$,
the summation of neutrino mass $\sum m_\mu$,
and the effective number of relativistic degrees of freedom in the Universe, $N_{\rm eff}$, via their degeneracies with $\Omega_m$ and $w$.
The early universe parameters, such as the running  $dn_s/d\ln k$ and the tensor-to-scalar ratio r, are less affected.

\begin{itemize}
 \item Due to the degeneracy in their roles of governing the cosmic distances,
 $\Omega_k$ is negatively correlated with $\Omega_m$ and positively correlated with $w$.
 As shown in Table \ref{Table:Exts},
 after considering AP the absolute value of $\Omega_k$ increases (0.28$\sigma$)
 changes the sign form minus to plus for the $\Lambda$CDM extension model.
 For wCDM extension model, the absolute value of $\Omega_k$ becomes bigger (0.14$\sigma$),
 but the error is more tightly constrained (32\% improvement).
\item Using the AP effect
the upper limit of the total neutrino mass are
both reduced (by 22\% and 18\%)
for $\Lambda$CDM extension and wCDM extension cases.
\item By comparing two scenarios of Planck+BAO and Planck+BAO+AP datasets,
$N_{\rm eff}$ is increased (0.6$\sigma$ for $\Lambda$CDM extension,
not obvious in case of $w$CDM extension).
The reduction in the error of $N_{\rm eff}$ is small
(from 11.4\% to 10.7\% for $\Lambda$CDM extension,
and from 12.7\% to 12.1\% for wCDM extension).
\item The running of the spectral index $dn_s/d\ln k$ is typically small,
while the error is not significantly changed in both cases.
\item Adding the AP test,
it appears that the tensor-to-scalar ratio $r$ is widely constrained for $\Lambda$CDM extension,
and has tighter constraints  for $\Lambda$CDM extension;
but considering the statistical significance (0.05$\sigma$ and 0.02$\sigma$),
the effect is really ignorable.
\end{itemize}

%
%
%
%
%

\begin{figure}
   \centering{
   \includegraphics[width=7cm]{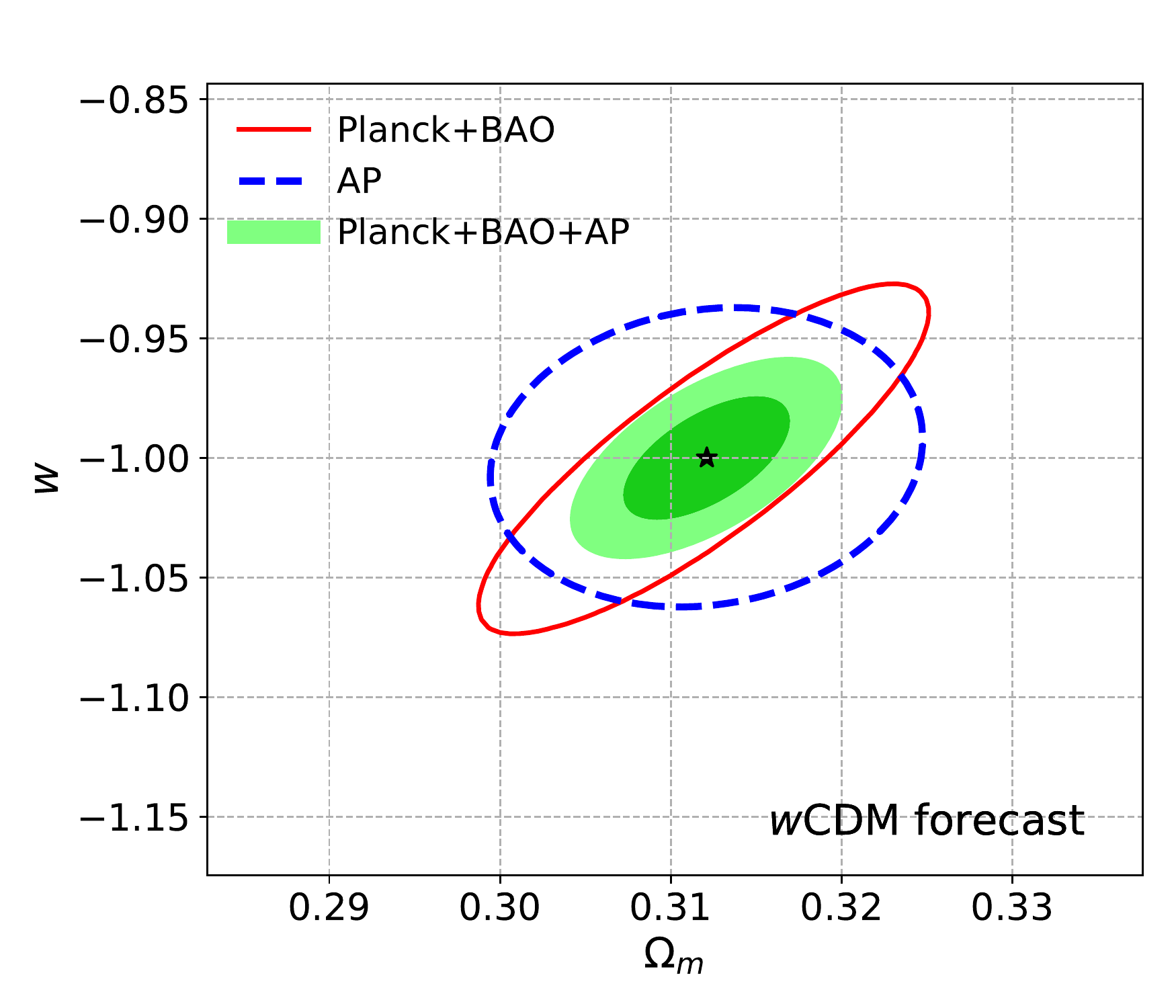}}
   \centering{
   \includegraphics[width=7cm]{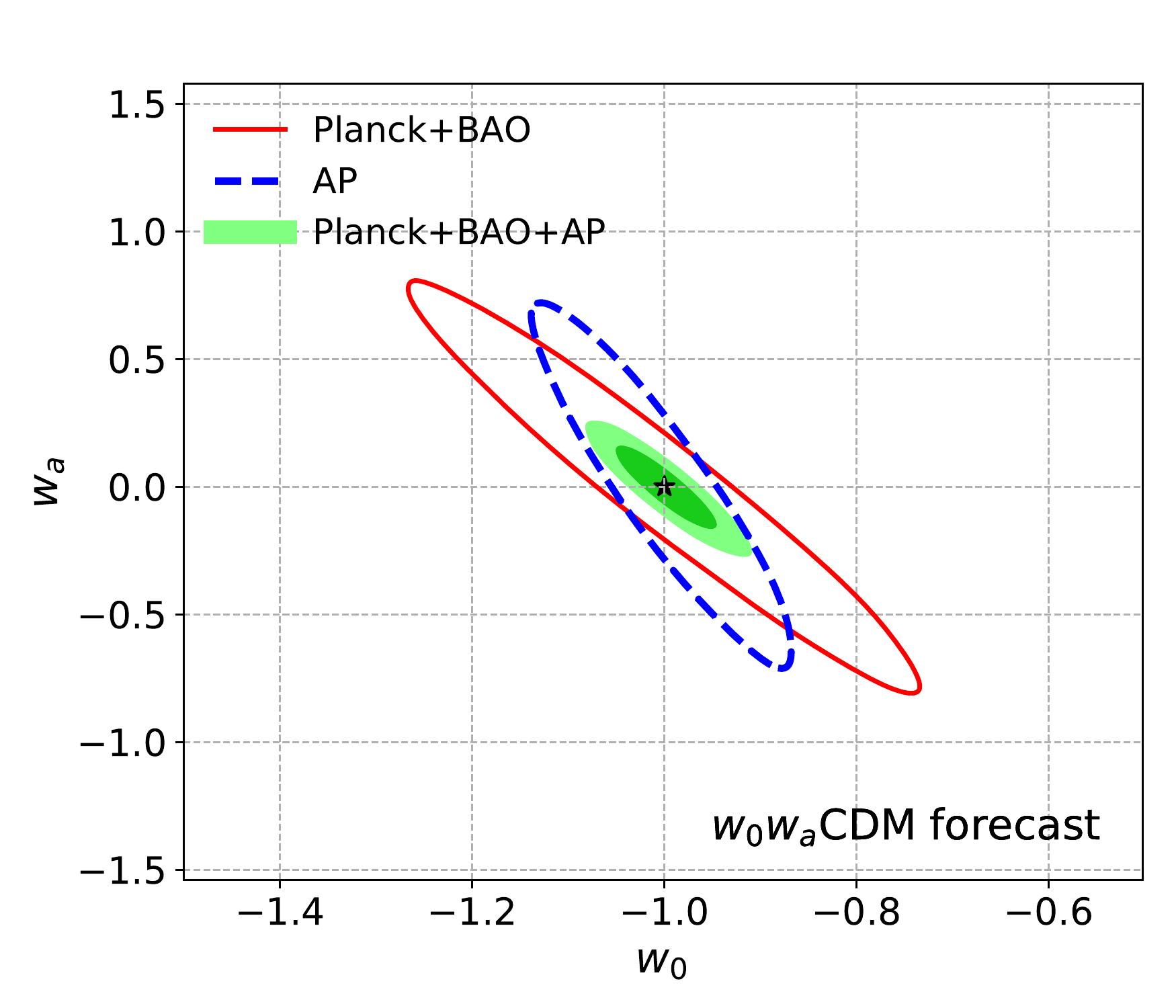}
    }
   \caption{\label{figfc}
   Expected constraints on $w$CDM and $w_0w_a$CDM models from DESI survey.
   Combining AP with Planck+BAO breaks the degeneracy between parameters,
    so reduces the uncertainty of $w$ by 50\%,
    and improving the $w_0$-$w_a$ constraint by a factor of 10.
   }
\end{figure}

The contour plots of one-parameter extensions to the $\Lambda$CDM model for combinations of Planck+BAO and Planck+BAO+AP
are illustrated in Figure \ref{figlcdmexts}.
We see clearly that combining the AP method increases the mean values of $N_{\mathrm{eff}}$, $\Omega_b h^2$, $n_s$, $H_0$,
decreases $\Omega_c h^2$, $\Omega_m$,
noticeably reduced the errors of $\Sigma m_{\nu}$.

The contour plots of one-parameter extensions to the $w$CDM model are illustrated in Figure \ref{figlcdmexts}.
In this case adding AP helps sharpens the constraints $\Omega_k$, $\sum m_{\nu}$, $H_0$, $\sigma_8$, $\Omega_\mathrm{m}$, $w$ by 20-30\%.


\section{Forecast }
\label{sec:FutureSurvey}

The number distribution of DESI galaxies is shown in Figure \ref{figngal}.
Predicted constraints are made for the $w$CDM and $w_0w_a$CDM.

Estimating the covariance matrix of our correlation-based estimator is a complicated job \citep{Bernstein:1993nb,OConnell:2015src}.
Among various terms that scale differently with $N_{gal}$, we found that
\begin{equation} \label{eq:covmat}
 {\bf Cov} \varpropto 1/N_{\rm gal},
\end{equation}
where $N_{\rm gal}$ is the number of galaxies,
is already a good approximation.
In Appendix \ref{sec:Appendixes:wCDM} we tested it using SDSS galaxies
and find its error is $\lesssim1.5\%-6\%$.

The DESI covariance matrices are then obtained simply using Eq. (\ref{eq:covmat}).
Firstly, the covariance matrix of $\hat\xi_{\Delta s}(z_2,z_1,\mu)$ using SDSS galaxies are computed using the 2,000 MD-PATCHY mocks.
We then choose this matrix as the baseline,
take a ratio of the $N_{gal}$s of SDSS and DESI,
and multiply this matrix by a factor to get the DESI covariance matrix
\footnote{ We take the SDSS galaxies in redshift bins 1,2 as the baseline to infer the DESI covariance matrices in all redshift bins.
The results are rather insensitive to the redshifts of the baseline galaxies.}.

The contour plot of $\Omega_m$ - $w$ in the wCDM model are illustrated at the top panel of Figure \ref{figfc}, using joint datasets of Planck+BAO,
AP and Planck+BAO+AP, respectively.
In the lower panels we show the contour plots of $w_0$ - $w_a$ in the frame of $w_0w_a$CDM model.
In all cases we find that the constraint greatly reduced after adding the AP method.

For the constraint on a single parameter, the performance of AP and Planck+BAO are comparable to each other.
If considering the joint constraint on two or more parameters,
then the different directions of degeneracy from the two sets of results suggests that
a greatly improved constraint can be achieved by combining them together.

In $w$CDM, adding AP reduces the constrained parameter space by 50\%,
achieving a precision of
\begin{equation}
 \delta \Omega_m\approx0.003,\ \delta w\approx0.015\ (\rm Planck+DESI\ BAO+DESI\ AP).
\end{equation}
In $w_0w_a$CDM, the addition of AP greatly reduces the constrained region by a factor of 10,
achieving a precision of
\begin{equation}
 \delta w_0\approx0.035,\ \delta w_a\approx0.11\ (\rm Planck+DESI\ BAO+DESI\ AP).
\end{equation}

\section{Concluding Remarks}
\label{sec:conclusion}

We conduct a comprehensive study about the cosmological constraints derived from tomographic AP method.
Based on \cite{Li2014,Li2015,Li2016,Li2018},
we improve the methodology by including the full covariance among clustering in all redshift bins.
We then apply it to current and future observational data.

When applying it to current observational data, we find:
\begin{itemize}
 \item The AP method noticeably improves the constraints on background evolution parameters $\Omega_m$, $H_0$, $w$, $w_0$, $w_a$.
 When combined it with the Planck+BAO the parameters' error bars are reduced by $\sim 20-50\%$ (depends on the model and parameter).
 \item Using Planck+BAO+SNIa+$H_0$+AP,
  a dynamical dark energy $w_a\neq -1$ is preferred at $\approx1.5 \sigma$ CL.
 \item In the framework of $\Lambda$CDM, adding AP into Planck+BAO yields to a slightly smaller  $\Omega_m=0.301\pm0.010$,
 and a slightly larger $H_0=68.9\pm1.2$.
 This leads to $\lesssim1\sigma$ changes in $\Omega_b h^2$, $n_s$, $z_*$, $r_*$, $z_{re}$.
 $\tau$, $A_s$, $\sigma_8$ are less affected.
 \item When considering 1-parameter extensions to $\Lambda$CDM and $w$CDM models,
 we get improved constraints on  $\Omega_k$, $\sum m_\mu$, $N_{\rm eff}$
 when combining AP with Planck+BAO.
 Since AP only puts constraints on the late time expansion,
 early universe parameters $dn_s/d{\rm ln}k$ and $r$ are less affected.
\end{itemize}

We make a forecast of the $w$CDM and $w_0w_a$CDM constraints expected from Planck+DESI.
We find the AP's constraints on $\Omega_m$, $w$, $w_0$ and $w_a$ are as tight as the Planck+BAO ones,
while the directions of degeneracy from the two differ from each other.
Thus, combining them significantly improves the power of constraint.
Adding AP reduces the error bar of $w$ by 50\%,
and improving the $w_0$-$w_a$ constraint by a factor of 10.

It should be pointed out that the many results presented in this work are {\it not} the optimistic ones.
We expect the result being further improved with the improvement in the methodology, e.g. optimistic binning scheme of the galaxies,
more aggressive clustering scales, more precise estimation of the covariance matrix, and so on.

According to our tests, for current surveys the systematic effects
can not significantly affect the derived cosmological constraints.
But it remains to be seen if this is true for future galaxy surveys.
In particular, the systematic effects are estimated using one set of simulation performed in a fiducial cosmology,
so the cosmological dependence of the systematics remains to be investigated in future works.
It could be solved by, e.g. interpolating among systematics estimated from several sets of simulations with different cosmologies,
considering theoretical estimation of systematics, and so on.




The tomographic AP method is so far the best method in separating the AP signal from the RSD distortions
and using it we already achieved strong cosmological constraints.
It is among the most powerful methods which can extract information from the $<$40 $h^{-1} \rm Mpc$ small-scale clustering region.
It is essentially important for us to improving this method and preparing for its application to the next generation surveys.


\section*{Appendix A: The Full Covmat Method Compared with the Old Method}
\label{sec:Appendix:FullCovmat}

As described in Sec. \ref{sec:methodology},
if one were ignoring the correlations between different $\delta_{i,j}\xi_{\Delta s}$s,
and simply using Eq. \ref{eq:chisq1} to calculate the $\chi^2$,
the result suffers from two problems:
\begin{itemize}
 \item The statistical power is over-estimated since part of the correlations are not considered;
 \item The result has a special dependence on the choice of the redshift bin (here the 1st bin),
 and thus suffers from large statistical fluctuation.
\end{itemize}

Here we conduct a simple test to see how large the above two effects are.
We simply consider six independent variables obeying normal distribution,
who share the same variance but have different mean values:
\begin{equation}
 X_i \sim \mathcal{N}(\mu=10 i,\sigma^2=1),\ i=1,2,3,4,5,6.
\end{equation}
We then use the ideas of Eq. (\ref{eq:chisq1},\ref{eq:chisq2})
to define a $\chi^2$ function characterizing the evolution among them:
\begin{eqnarray}
 \chi^2_1 &\equiv& \sum_{i=2}^6(X_i-X_1)^2/(\sigma^2+\sigma^2),\\
 \chi^2_2 &\equiv& \sum_{i=2}^6 \sum_{j=2}^6 (X_i-X_1)\ {\bf Cov}_{i,j}\ (X_j-X_1),
\end{eqnarray}
and here the covariance matrix simply takes the form of
\begin{equation}
 {\bf Cov}_{ii} = 2;\ {\bf Cov}_{ij}=1\ \rm for\ i\neq j.
\end{equation}

\begin{figure}
	\centering{
		\includegraphics[height=7cm]{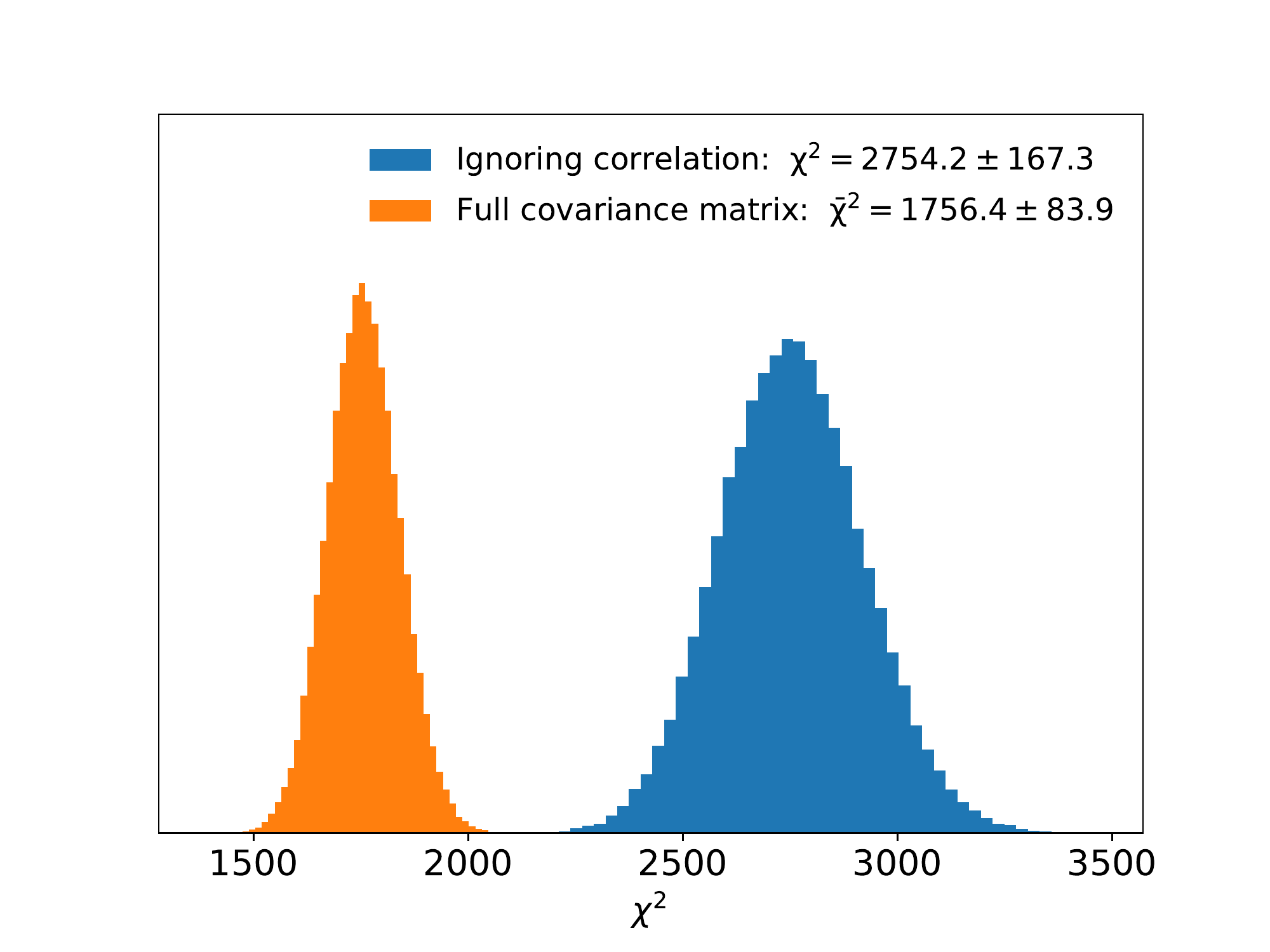}
	}
	\caption{\label{figa1}
		Histogram of $\chi^2$ values using six test variables as described in Section \ref{sec:Appendix:FullCovmat}.
		The full covariance matrix method results in reliable and stable estimation of $\chi^2$s,
		while ignoring the covariance overestimates the $\chi^2$s and suffers from significantly larger uncertainty.
	}
\end{figure}

We generate $10^6$ sets of $X_i$, compute their $\chi^2$ values, and plot the result in Figure \ref{figa1}.
The mean and root-mean-square are listed in the legend.
We find that
\begin{itemize}
 \item In this case, the part-cov approach overestimates the $\chi^2$
 value by 57\%;
 i.e., it over estimates the statistical significance
 of the evolution.
 \item The statistical fluctuation of the $\chi^2$ derived from the
 part-cov approach is twice as large as the full-cov approach.
 This may lead to large bias when adopting this method to constrain cosmological parameters.
\end{itemize}

\begin{figure}
   \centering{
   \includegraphics[width=7cm]{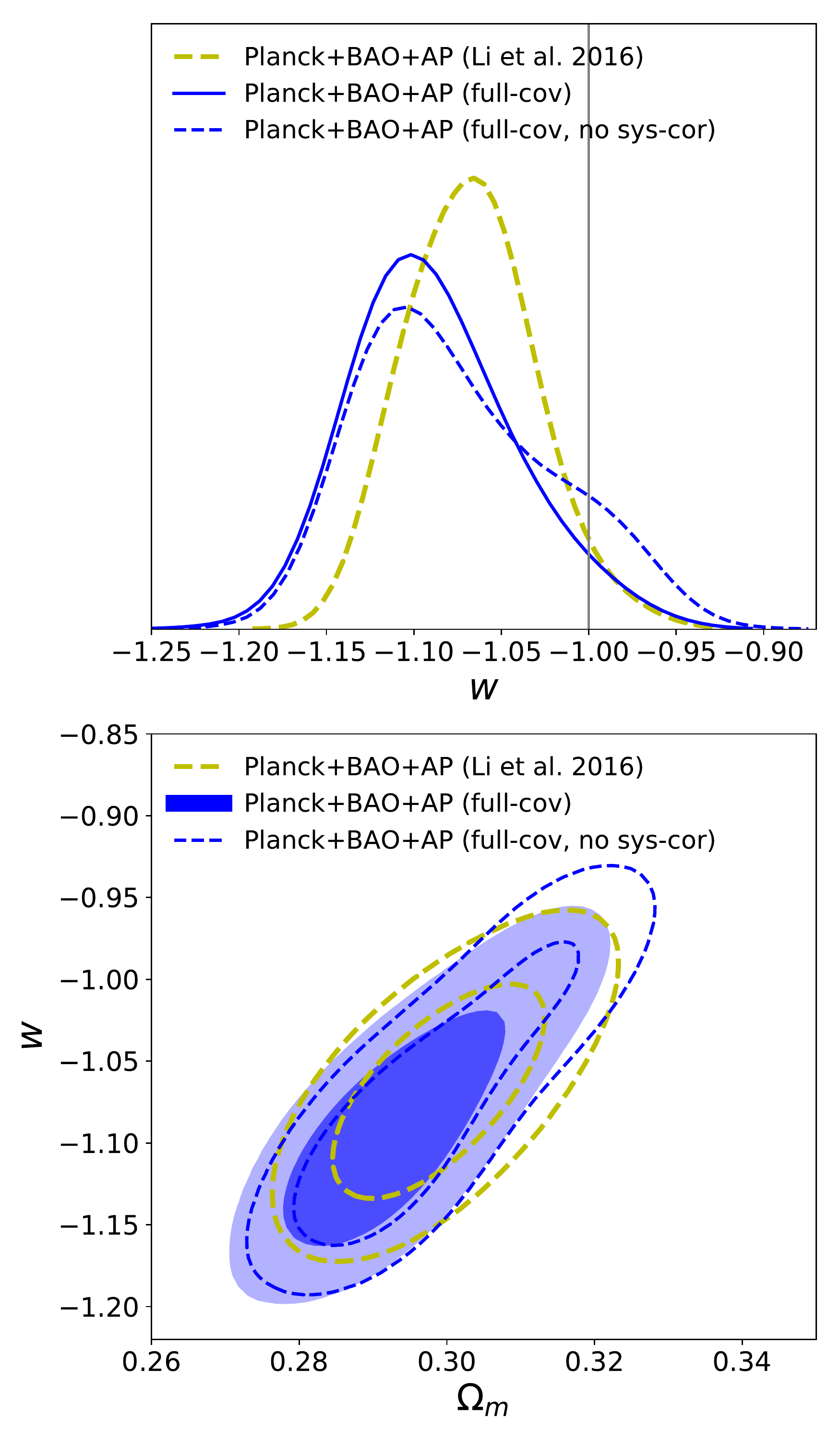}
    }
   \caption{\label{figwcdmtest}
   $w$CDM.
   1) The method we used in this paper (full covmat method) is different from Li et al. 2016 (where they do not use full covmat).
   Result is slightly different.
   2) If we discard systematic correction (estimated from HR4 simulation), there is minor change in the results.
   }
\end{figure}

Figure \ref{figwcdmtest}
compares the Planck+BAO+AP constraint on $w$CDM
derived using the 1st-bin approach Eq.
(yellow dashed line)
and the full-covmat approach (blue line).
We find
\begin{eqnarray}\label{eq:part_full}
 && w= -1.068\ ^{+0.036}_{-0.040}\ ^{+0.076}_{-0.073}\ {\rm par-cov},\\
 && w= -1.089\ ^{+0.040}_{-0.054}\ ^{+0.104}_{-0.088}\ {\rm full-cov}.
\end{eqnarray}
When using the full-covmat approach,
the mean value was shifted towards the negative values by 0.02 ($\sim 0.4 \sigma$),
and the upper/lower error bars are enlarged by 10\%/26\%, respectively
The two sets of results are, still, in statistical consistency.
The errors due to the defect of the 1st-bin approach are not  serious.

The part-cov constraint is weaker than what reported in \cite{Li2015} ($w=-1.054\pm0.025$)
because the difference in the choices of $n_\mu$ (\citep{Li2015} adopted $n_\mu=6-40$ and here we reduced it to 20-25).
Different from \cite{Li2015}, in this work,
we adopt the technique developed in \cite{Li2018} to
efficiently approximate the 2PCFs in different cosmologies,
and increase the size of covariance matrix from
$n_\mu \times n_\mu$ to $5n_\mu \times 5n_\mu$.
Both changes make the analysis more sensitive to the noise in the 2PCFS.
So we reduce the number of binning to
reduce the noise in $\xi_{\Delta s}$,
which increase the reliability of the results,
in the cost of scarifying some power of constraints.

\begin{figure}
   \centering{
   \includegraphics[width=7cm]{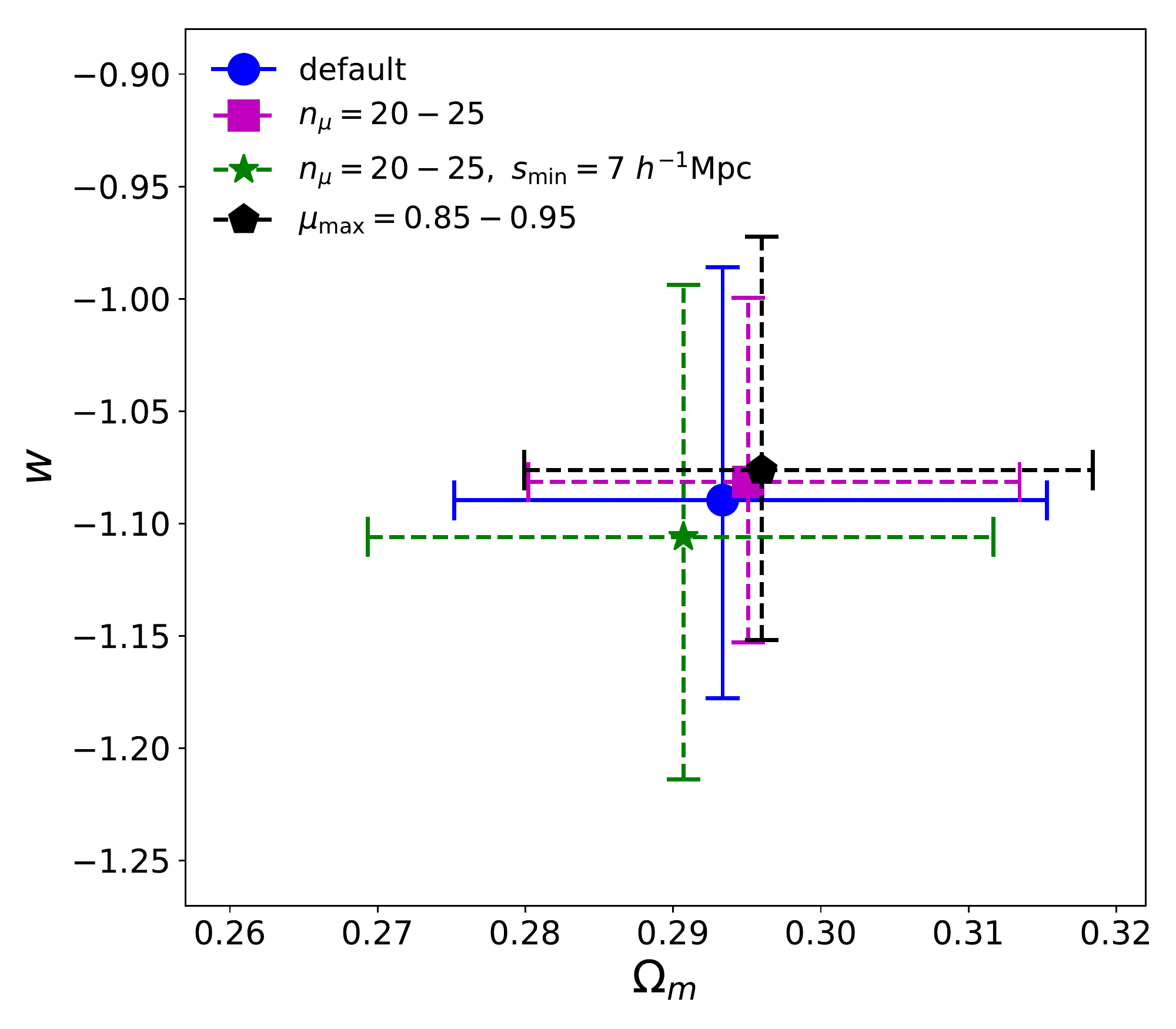}
    }
   \caption{\label{figwcdmtest1}
   $w$CDM.
  Plotted are 2$\sigma$ regions of parameters. If we change the options in the analysis, changes in the results are minor.
   }
\end{figure}

\begin{figure}
   \centering{
   \includegraphics[width=9cm]{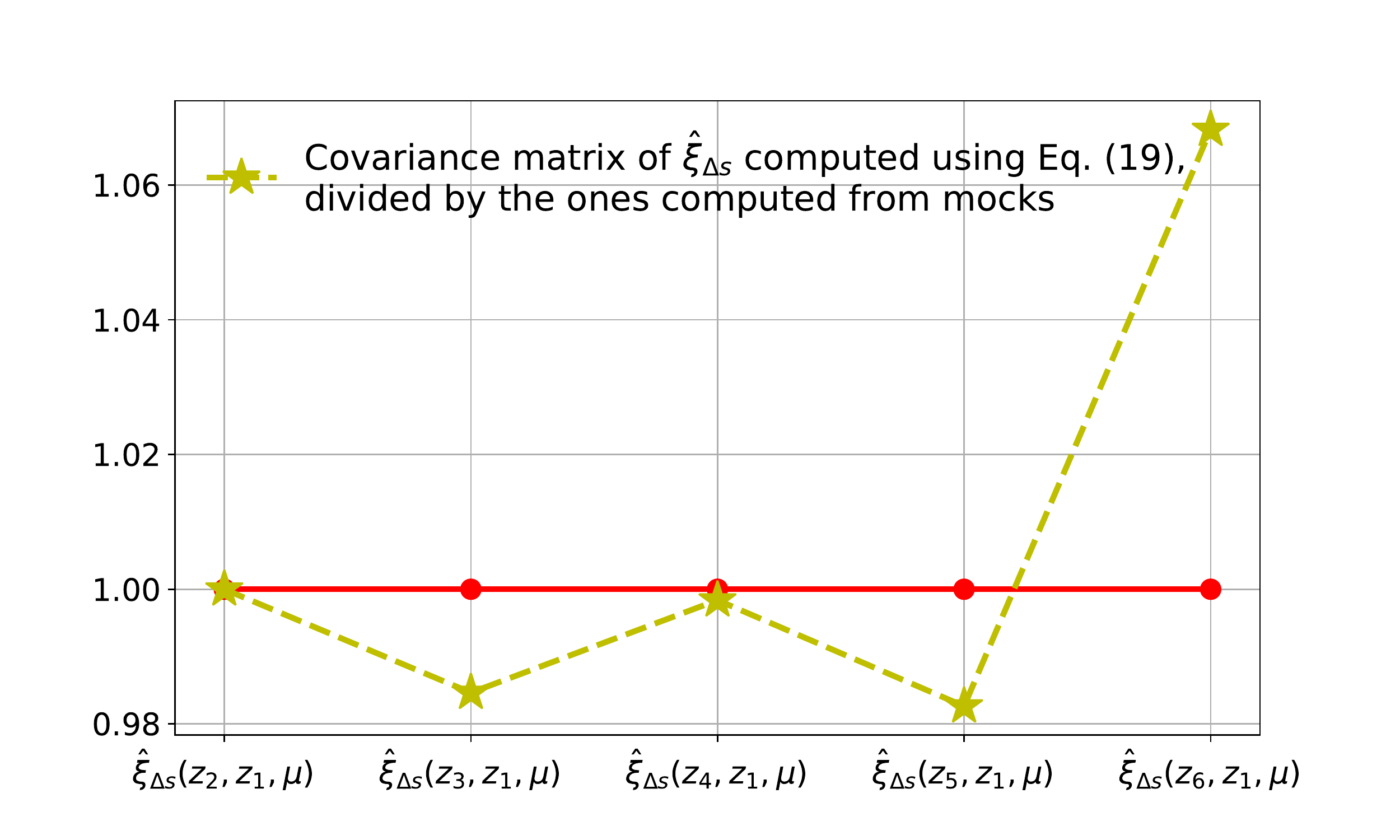}
    }
   \caption{\label{fig_sigma}
A rough check of Eq. (\ref{eq:covmat}) using 6 bins of SDSS DR12 galaxies (see Figure 2 of Li16).
The covariance matrices of $\hat\xi_{\Delta s}(z_i,z_1,\mu)$, $i=1,2,3,4,5$ are computed using two methods:
1. Measuring the scattering of $\delta_{i,j}$ from the 2,000 MultiDark-Patchy mocks, as we did in \cite{Li2016}.
2. Measuring $\delta_{2,1}$ from the mocks, and infer the other $\delta_{i,j}$s simply using Eq. (\ref{eq:covmat}).
The y-axis shows the ratio between the results of the two methods (method 2 over method 1).
We find the estimation (method 2) achieve $<2\%$ precision for $\delta_{3,1}$, $\delta_{4,1}$, $\delta_{5,1}$.
In the related redshift bins (redshift bins 1-5), the galaxy number density $\bar n_{\rm gal}$ varies in the range of $2-7 \times 10^{-4} h^{-1}$ Mpc
(as large as $\sim$3 times).
Given such a large fluctuation of $\bar n_{\rm gal}$, Eq. (\ref{eq:covmat}) still achieves good precision.
In the sixth bin $\bar n_{\rm gal}$ drops to significantly lower density ($1 \times 10^{-4} (h^{-1} \rm Mpc)^{-3}$,
$\sim$5 times lower than the first and second bins), while the error of method 2 is still $\lesssim$6\%.
   }
\end{figure}

\section*{Appendix B: Robustness Check}
\label{sec:Appendixes:wCDM}

Li16 tested the robustness of the tomographic AP method in details,
and found the derived constraints on $w$CDM are insensitive to the adopted options within the range of $s_{\rm min}=2-8\ h^{-1}$ Mpc, $s_{\rm max}=30-50\ h^{-1}$ Mpc, $\mu_{\rm max}=0.85-0.99$,
and number of binning $n_\mu=6-40$.
\cite{Li2018} re-conducted the above tests in the $w_0 w_a$CDM and obtained similar result.
In what follows, we adopt the procedure of these two papers and test the robustness of the result.
Our default set of options are $s_{\rm min}=6$  $h^{-1}$ Mpc,
$s_{\rm max}=40\ h^{-1}$ Mpc, $\mu_{\rm max}=0.97$, and $n_\mu=15-20$
\footnote{Our $n_\mu$ is smaller than the default choice of Li16 (where they more ambitiously chose $n_\mu=20-35$);
this weakens the constraints by a little bit, but increases the robustness of the results,
and also reduces the noise in the likelihood.}.

In the two panels of Figure \ref{figwcdmtest} we also plotted the results without conduction any systematics (blue dashed).
We see the peak value of $w$ remain almost unchanged;
the only effect is a small ($\sim 10\%$ for the 95\% contour) enlargement of constrained region towards the
larger value of $w$ and $\Omega_m$.
This is similar to what we found in \cite{Li2018},
i.e. for the analysis of current observations the systematics effect
is not large enough to result in a statistically significant change of the results.

As a detailed test of the options adopted in the analysis,
Figure \ref{figwcdmtest1} shows the mean values and 95\% limits of the parameters,
in cases of using a more aggressive binning scheme $n_\mu=20-25$,
a more conservative small-scale cut $s_{\rm 7}= h^-1$Mpc,
and a more conservative cut of correlation angle $\mu_{\rm max=0.85-0.95}$.
In all cases we find rather small change in the mean values ($\lesssim0.2\sigma$)
and the limits ($\lesssim15\%$).
So we conclude the cosmological constraints obtained from the AP method does not sensitively depend on these options.

Finally, we test precision of Eq. \ref{eq:covmat} on the BOSS galaxies,
and found it works with satisfying precision (Figure \ref{fig_sigma}).
The difference between the covariance matrices estimated from Eq. \ref{eq:covmat} and those directly computed using the mocks is $\lesssim$5\%.

\section*{Acknowledgments}

We acknowledges the usage of Horizon Run simulations and MultiDark-Patchy mocks.
We thank Zhiqi Huang and Xin Zhang for helpful discussions.
XDL gratefully acknowledge professor Changbom Park,
whose insights and persistence makes the novel method becomes reality.
XDL acknowledges the supported from NSFC grant (No. 11803094).
QGH is supported by grants from NSFC
(grant No. 11335012, 11575271, 11690021, 11747601),
the Strategic Priority Research Program of Chinese Academy of Sciences
(Grant No. XDB23000000), Top-Notch Young Talents Program of China,
and Key Research Program of Frontier Sciences of CAS.

Based on observations obtained with Planck (http://www.esa.int/Planck),
an ESA science mission with instruments and contributions directly funded by
ESA Member States, NASA, and Canada.

Funding for SDSS-III has been provided by the Alfred P. Sloan Foundation, the Participating Institutions, the
National Science Foundation, and the U.S. Department of Energy Office of Science.
The SDSS-III web site is http://www.sdss3.org/.
SDSS-III is managed by the Astrophysical Research Consortium for the Participating Institutions
of the SDSS-III Collaboration including the University of Arizona, the Brazilian Participation Group, Brookhaven
National Laboratory, Carnegie Mellon University, University of Florida, the French Participation Group,
the German Participation Group, Harvard University, the Instituto de Astrofisica de Canarias, the Michigan State/Notre
Dame/JINA Participation Group, Johns Hopkins University, Lawrence Berkeley National Laboratory, Max Planck
Institute for Astrophysics, Max Planck Institute for Extraterrestrial Physics, New Mexico State University, New
York University, Ohio State University, Pennsylvania State
University, University of Portsmouth, Princeton University,
the Spanish Participation Group, University of Tokyo, University of Utah, Vanderbilt University, University of Virginia,
University of Washington, and Yale University.


\label{lastpage}

\end{document}